\documentclass[journal,11pt,draftclsnofoot,onecolumn]{IEEEtran}

\usepackage{latexsym}
\usepackage{amsfonts}
\usepackage{graphicx} 
\usepackage{subfigure}

\usepackage{amssymb}
\usepackage{natbib}
\usepackage{enumerate}
\usepackage{amsmath}

\usepackage{tabularx}

\usepackage[usenames]{color}

\newcommand{\dfn}{\triangleq}

\def\BibTeX{{\rm B\kern-.05em{\sc i\kern-.025em b}\kern-.08em
    T\kern-.1667em\lower.7ex\hbox{E}\kern-.125emX}}

\newcolumntype{Z}{>{\raggedleft\arraybackslash}X}

\begin{document}

%
\title{Improved Adaptive Rejection Metropolis Sampling Algorithms}
\author{Luca Martino$^\dagger$, Jesse Read$^\dagger$, David Luengo$^\ddagger$\\
$^\dagger$Department of Signal Theory and Communications, Universidad Carlos III de Madrid.\\
Avenida de la Universidad 30, 28911 Legan\'es, Madrid, Spain.\\
$^\ddagger$Department of Circuits and Systems Engineering, Universidad Polit\'ecnica de Madrid.\\
Carretera de Valencia Km. 7, 28031 Madrid, Spain.\\
E-mail: {\tt luca@tsc.uc3m.es, jread@tsc.uc3m.es, david.luengo@upm.es}}
\maketitle 

\vspace{-1.5cm}

\begin{abstract}
Markov Chain Monte Carlo (MCMC) methods, such as the Metropolis-Hastings (MH) algorithm, are widely used for Bayesian inference.
One of the most important challenges for any MCMC method is speeding up the convergence of the Markov chain, which depends crucially on a suitable choice of the proposal density.
Adaptive Rejection Metropolis Sampling (ARMS) is a well-known MH scheme that generates samples from one-dimensional target densities making use of adaptive piecewise linear proposals constructed using support points taken from rejected samples.
%
%
In this work we pinpoint a crucial drawback of the adaptive procedure used in ARMS: support points might never be added inside regions where the proposal is below the target.
When this happens in many regions it leads to a poor performance of ARMS, and the sequence of proposals never converges to the target.
In order to overcome this limitation, we propose two alternative adaptive schemes that guarantee convergence to the target distribution.
These two new schemes improve the adaptive strategy of ARMS, thus allowing us to simplify the construction of the sequence of  proposals.
%
%
%
%
Numerical results show that the new algorithms outperform the standard ARMS and other techniques.
\end{abstract}


\IEEEpeerreviewmaketitle

\section{Introduction}









Bayesian inference techniques and their implementation by means of sophisticated Monte Carlo (MC) statistical methods, such as Markov chain Monte Carlo (MCMC) and sequential Monte Carlo (SMC) approaches (also known as particle filters), has become a very active area of research over the last years \citep{Gilks95bo,Gamerman97bo,Liu04b,Robert04,Liang10}.
Monte Carlo techniques are very powerful tools for numerical inference, stochastic optimization and simulation of complex systems \citep{Devroye86,Fearnhead98,Doucet01b,Jaeckel02,Robert04}.

Rejection sampling (RS) \citep{Devroye86,VonNeumann51} and the Metropolis-Hastings (MH) algorithm are two well-known classical Monte Carlo methods for universal sampling \citep{Metropolis49,Metropolis53,Hastings70,Liu04b}.
Indeed, they can be used to generate samples (independent with the RS and correlated with MH) from virtually any target probability density function (pdf) by drawing from a simpler proposal pdf.
Consequently, these two methods have been widely diffused and applied. 
Unfortunately, in some situations they present certain important limitations and drawbacks as we describe below.

In the RS technique the generated samples are either accepted or rejected by an adequate test of the ratio of the target, $p(x)$, and the proposal density, $\pi(x)$, with $x \in \mathcal{D} \subseteq \mathbb{R}$.\footnote{Note that we assume, as is commonly done in the literature, that both pdfs are known up to a normalization constant, and we denote as $p_o(x)\propto p(x)$ and $\tilde{\pi}(x)\propto \pi(x)$ the normalized target and proposal pdfs.}
An important limitation of RS is the need to analytically establish an upper bound $M$ for the ratio of these two pdfs, $M\geq \frac{p(x)}{\pi(x)}$ or equivalently $M\pi(x)\geq p(x)$, which is not always an easy task.
Moreover, even using the best bound, i.e. $M=\sup_x p(x)/\pi(x)$, the acceptance rate can be very low if the proposal is not similar to the target pdf.
On the oher hand, in the MH algorithm, depending on the choice of the proposal pdf, the correlation among the samples in the Markov chain can be very high \citep{Liu04b,Liang10,MartinoSigPro10}.
Correlated samples provide less statistical information and the resulting chain can remain trapped almost indefinitely in a local mode, meaning that convergence will be very slow in practice.
Moreover, determining how long the chain needs to be run in order to converge is a difficult task, leading to the common practice of establishing a large enough burn-in period to ensure the chain's convergence \citep{Gilks95bo,Gamerman97bo,Liu04b,Liang10}. 

In order to overcome these limitations several extensions of both techniques have been introduced \citep{Devroye86,Liu04b,Liang10}, as well as methods combining them \citep{Tierney91,Tierney94}.
Furthermore, in the literature there is great interest in adaptive MCMC approaches that can speed up the convergence of the Markov chain \citep{Haario01,Gasemyr03,Andrieu06,Andrieu08,Holden09,Griffin11}.
Here we focus on two of the most popular adaptive extensions, pointing out their important limitations, and proposing a novel adaptive technique that can overcome them and leads to a much more efficient approach for drawing samples from the target pdf.

One widely known extension of RS is the class of {\em adaptive rejection sampling} (ARS) methods  \citep{Gilks92derfree,Gilks92,CorrGilks97}, which is an improvement of the standard RS technique that ensures high acceptance rates with a moderate and bounded computational cost.
Indeed, the standard ARS algorithm of \citep{Gilks92} yields a sequence of proposal functions, $\{\pi_t(x)\}_{t\in \mathbb{N}}$, that converge towards the target pdf when the procedure is iterated.
The basic mechanism of ARS is the following.
Given a set of support points, $\mathcal{S}_t=\{s_1,...,s_{m_t}\}$, ARS builds a proposal pdf composed of truncated exponential pdfs inside the intervals $(s_j,s_{j+1}]$ ($0 \le j \le m_t$), considering also the open intervals $(-\infty, s_1]$ (corresponding to $j=0$) and $[s_{m_t},+\infty)$ (associated to $j=m_t$) when $\mathcal{D}= \mathbb{R}$.
Then, when a newly proposed sample $x'$, drawn from $\pi_t(x)$, is rejected, it is always added to the set of support points, $\mathcal{S}_{t+1}=\mathcal{S}_{t}\cup \{x'\}$, which are used to build a refined proposal pdf, $\pi_{t+1}(x)$, for the next iteration.

Note that the proposal density used in ARS becomes quickly closer to the target pdf and the proportion of accepted samples grows.
Consequently, since the proposal pdf is only updated when a sample is rejected and the probability of discarding a sample decreases quickly to zero, the computational cost is bounded and remains moderate, i.e. the computational cost does not diverge due to this smart adaptive strategy that improves the proposal pdf only when and where it is needed.
Unfortunately, this algorithm can only be used with log-concave target densities and hence also unimodal. Therefore, several generalizations have been proposed \citep{Gilks95,Hoermann95,Evans98,Gorur08rev,MartinoStatCo10} handling specific classes of target distributions or using jointly a MCMC approach.

Indeed, for instance, another popular technique that combines the ARS and MH approaches in an attempt to overcome the limitations of both methodologies is
the {\em Adaptive Rejection Metropolis Sampling} (ARMS) \citep{Gilks95}.
ARMS extends ARS to tackle multimodal and non-log-concave densities by allowing the proposal to remain below the target in some regions and adding a Metropolis-Hastings (MH) step to the algorithm to ensure that the accepted samples are properly distributed according to the target pdf.
Note that introducing the MH step means that, unlike in the original ARS method, the resulting samples are correlated.
The ARMS technique uses first an RS test on each generated sample $x'$ and, if this sample is initially accepted, applies also an MH step to determine whether it is finally accepted or not.
If a sample $x'$ is rejected in the RS test (hence, imperatively $\pi_t(x') \geq p(x')$, as otherwise samples are always initially accepted by the RS step), then it is incorporated to the set of support points, $\mathcal{S}_{t+1}=\mathcal{S}_{t}\cup \{x'\}$, that is used to build a better proposal pdf for the next iteration, $\pi_{t+1}(x)$, exactly as in ARS.
On the other hand, when a sample is initially accepted, after the MH step the set of support points is never updated, i.e. $\mathcal{S}_{t+1}=\mathcal{S}_{t}$, meaning that the proposal pdf is not improved and $\pi_{t+1}(x)=\pi_t(x)$.

This is the crucial point w.r.t. the performance of ARMS.
In general, if proper procedures for constructing $\pi_t(x)$ given a set of support points, $\mathcal{S}_{t}$, such as the ones proposed in \citep{Gilks95,Meyer08} are adopted, the proposal will improve throughout the whole domain $x\in \mathcal{D}$.
However, inside the regions of the domain $\mathcal{D}$ where $\pi_t(x) < p(x)$, new support points might never be added.
Consequently, the convergence of the sequence $\{\pi_t(x)\}_{t\in \mathbb{N}}$ to $p(x)$ cannot be guaranteed, and it may not occur in some cases, as shown in Section \ref{SectStrucLim} through a simple example.
Due to this problem and the correlation among the samples generated, when the target pdf is multimodal the Markov chain generated by the ARMS algorithm tends to get trapped in a single mode despite the (partial) adaptation of the proposal pdf (see e.g. example 2 in \citep{MartinoSigPro10}). 

It is important to remark that this drawback is caused by a structural problem of ARMS: the adaptive mechanism proposed for updating the proposal pdf is not complete, since it never adds support points inside regions where $\pi_t(x) < p(x)$.
Therefore, even though a suitable procedure for constructing the proposals (see e.g. \citep{Gilks95,Meyer08}) can allow them to change inside these regions (eventually obtaining $\pi_t(x) \ge p(x)$ in some cases), the convergence cannot be guaranteed regardless of the procedure used to build the proposal densities.
For this reason, the performance of the standard ARMS depends critically on the choice of a very good way to construct the proposal pdf, $\pi_t(x)$, that attains $\pi_t(x') \geq p(x')$ almost everywhere, implying that ARMS tends to be reduced to the ARS algorithm.
Indeed, note that if the procedure used to build the proposal produces $\pi_0(x) < p(x)$ $\forall x\in \mathcal{D}$, then the proposal is never improved, i.e.  $\pi_t(x) = \pi_0(x)$ for all $t\in \mathbb{N}$, resulting in no adaptation at all.

This structural problem of the ARMS approach can be solved in a trivial way: adding a new support point each time that the MH step is applied.
Unfortunately, in this case the computational cost of the algorithm increases rapidly as the number of support points diverges.
Moreover, a second major problem is that the convergence of the Markov chain to the invariant density cannot be ensured, as partially discussed in \citep{Gilks95} and in \citep{CorrGilks97} for the case of ARMS-within-Gibbs (see also \cite[Chapter 8]{Liang10} for further considerations).

In this work we present two enhancements of ARMS that guarantee the convergence of the sequence of proposal densities to the target pdf with a bounded computational cost, as in the standard ARS and ARMS approaches, since the probability of incorporating a new support point quickly decreases to zero.
Moreover, the two novel techniques fulfill one of the required condition to assure the convergence of the chain, the so-called {\it diminishing adaptation} (see e.g. \cite[Chapter 8]{Liang10}, \citep{Haario01,Roberts07}).
The first one is a direct modification of the ARMS procedure that allows us to incorporate support points inside regions where the proposal is below the target with a decreasing probability as the proposal converges to the target.
The second one is an adaptive independent MH algorithm that learns from all past generated samples except for the current state of the chain, thus also guaranteeing the convergence of the chain to the invariant density, as shown in \citep{Gasemyr03,Holden09}. 
Furthermore, these new strategies also allow us to reduce the complexity in the construction of the proposal pdfs (thus reducing both the effort of writing the code and the computational cost of the resulting algorithm), since they do not require that $\pi_t(x) \ge p(x)$ almost everywhere as in the standard ARMS algorithm in order to guarantee the smooth running of the adaptive mechanism and hence to improve the proposal pdf everywhere in the whole  domain $\mathcal{D}$.
We exemplify this point introducing simpler procedures to construct the sequence of proposal densities and illustrating their good performance through numerical examples.

We remark that, compared to other adaptive Metropolis-Hastings techniques available in literature \citep{Warnes01,Haario01,Haario06,Cai08,Liang10}, the two new schemes provide a better performance.
For instance, two alternative adaptive schemes used to draw samples from one-dimensional target pdfs are the {\it adaptive triangle Metropolis sampling} (ATRIMS) and the {\it adaptive trapezoid Metropolis sampling} (ATRAMS) proposed in \citep{Cai08}.
However, even though the proposal pdfs are adaptively improved in both of them (ATRIMS and ATRAMS), the sequence of proposals does not converge to the target density, unlike the adaptive strategies proposed in this work. 
Regarding other adaptive MH algorithms \citep{Warnes01,Haario01,Haario06,Keith08,Hanson11} only certain parameters of the proposal pdf (which follows a parametric model with a fixed analytical form) are adjusted and optimized, whereas here we improve the entire shape of the proposal density, which becomes closer and closer to the shape of the target density.
Finally, another advantage of our schemes is that the proposed algorithms eventually become standard ARS techniques when the constructed proposal pdf is always above the target, i.e. when $\pi_t(x) \geq p(x)\ \forall x\in \mathcal{D}$, thus producing independent samples with acceptance rate that quickly becomes close to one.
For all these reasons, drawing from univariate pdfs, our techniques obtain better performance than the other techniques available in literature at the expense of a moderate increase in the computational cost. 

The rest of the paper is organized as follows. Background material is presented in Section \ref{BackSect}. Then we review and discuss certain limitations of ARMS in Sections \ref{Sect_ARMS} and \ref{SectStrucLim}. In Section \ref{Variants_Sect} we present the two novel techniques, whereas alternative procedures to construct the proposal pdf are described in Section \ref{SecOtherProc}. Finally, numerical simulations are shown in Section \ref{Sect_SIMU} and conclusions are drawn in Section \ref{conclusion_Sect}.

\section{Background}
\label{BackSect}
\subsection{Notation}
We indicate random variables with upper-case letters, e.g. $X$, while we use lower-case letters to denote the corresponding realizations, e.g. $x$.  
%
Sets are denoted with calligraphic upper-case letters, e.g. $\mathcal{R}$. The domain of the variable of interest, $x$, is denoted as $\mathcal{D} \subseteq \mathbb{R}$.  
The normalized target PDF is indicated as $p_o(x)$, whereas $p(x) = c_p p_o(x)$ represents the unnormalized target.
The normalized proposal PDF is denoted as $\tilde{\pi}(x)$, whereas $\pi(x) = c_{\pi} \tilde{\pi}(x)$ is the unnormalized proposal.
For simplicity, we also refer to the unnormalized functions $p(x)$ and $\pi(x)$, and in general to all unnormalized but proper PDFs, as densities.

\subsection{Adaptive rejection sampling}

The standard adaptive rejection sampling (ARS) algorithm  \citep{Gilks92derfree,Gilks92,CorrGilks97} enables the construction of a sequence of proposal densities, $\left\{\pi_t(x)\right\}_{t\in \mathbb{N}}$, tailored to the target density, $p_o(x)\propto p(x)$.
Its most appealing feature is that each time that a sample is drawn from a proposal, $\pi_t(x)$, and it is rejected, this sample can be used to build an improved proposal, $\pi_{t+1}(x)$, with a higher mean acceptance rate.
Unfortunately, the ARS method can only be applied with target pdfs which are log-concave (and thus unimodal), which is a very stringent constraint for many practical applications \citep{Gilks95,MartinoStatCo10}.
In this section we briefly review the ARS algorithm, which is the basis for the ARMS method and the subsequent improvements proposed in this paper.

\begin{figure}[htb]
\centering 
\subfigure[]{\includegraphics[width=6cm]{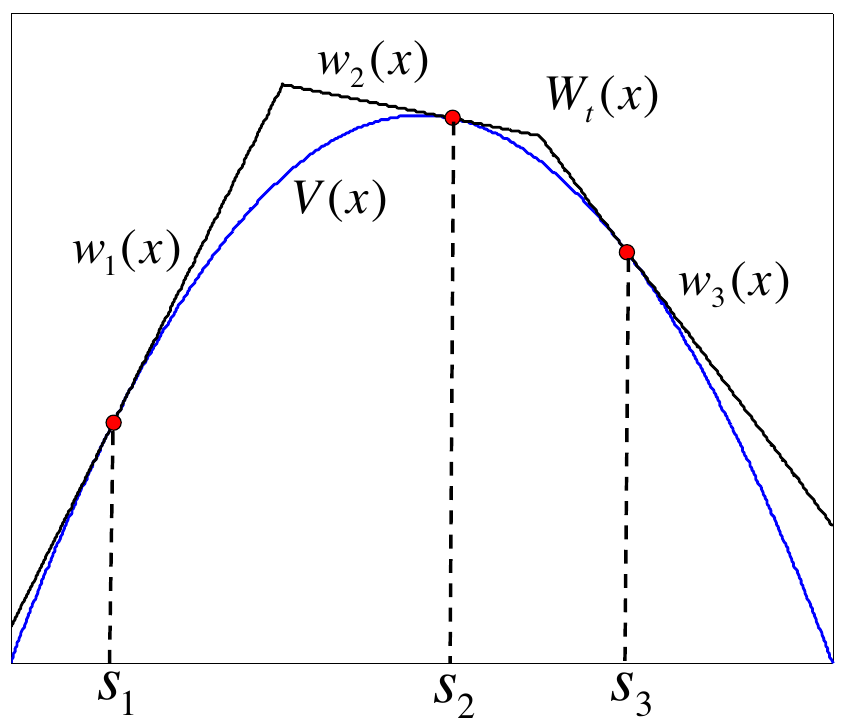}}
 \subfigure[]{ \includegraphics[width=6cm]{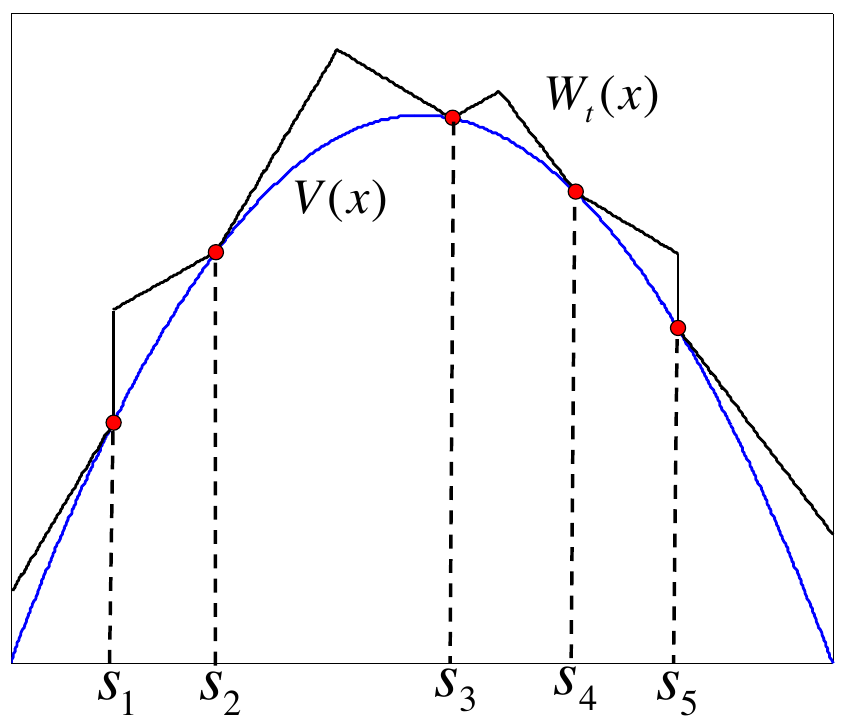}}
  \caption{Two possible procedures to build a suitable piecewise linear function $W_t(x)\geq V(x)$ when $V(x)$ is concave: {\bf (a)} using the first derivative of $V(x)$ (i.e. tangent lines) at the support points (with $3$ support points, for instance); {\bf (b)} using secant lines (with $5$ support points, for instance).}
\label{figARStwoProc}
\end{figure}

Let us assume that we want to draw samples from a target pdf, $p_o(x) \propto p(x)$, with support  $\mathcal{D} \subseteq \mathbb{R}$, known up a normalization constant.
The ARS procedure can be applied when $p(x)$ is log-concave, i.e. when
\begin{equation}
V(x)\dfn \log(p(x)),	
\end{equation}
is strictly concave $\forall x \in \mathcal{D} \subseteq \mathbb{R}$. 
In this case, let  
\begin{equation}
\mathcal{S}_t\dfn\{s_1, s_2,\ldots, s_{m_{t}}\}\subset \mathcal{D}
\end{equation}
be the set of {\it support points} at time $t$, sorted in ascending order (i.e. $s_1 < \ldots < s_{m_{t}}$). Note that the number of points $m_{t}$ can grow with the iteration index $t$. 
From $\mathcal{S}_t$ a piecewise-linear function $W_t(x)$ is constructed such that 
\begin{equation}
W_t(x)\geq V(x),
\end{equation}
for all $x\in\mathcal{D}$ and $\forall t\in\mathbb{N}$. Different approaches are possible for constructing this function.
For instance, if we denote by $w_{k}(x)$ the linear function tangent to $V(x)$ at $s_k\in \mathcal{S}_t$ \citep{Gilks92}, then a suitable piecewise linear function $W_{t}(x)$ is:
\begin{equation}
\label{ARSwithDer}
	W_{t}(x)\dfn\min\{w_{1}(x),\ldots,w_{m_{t}}(x)\} \mbox{  } \mbox{  for  } \mbox{  } x\in \mathcal{D}.
\end{equation}
In this case, clearly we have $W_t(x)\geq V(x)$, $\forall x\in \mathcal{D}$ and $\forall t\in\mathbb{N}$, as shown in Figure \ref{figARStwoProc}(a).
However, it is also possible to construct $W_{t}(x)$ without using the first derivative of $V(x)$, which is involved in the calculation of the tangent lines used in the previous approach, by using secant lines \citep{Gilks92derfree}.
Indeed, defining the intervals
\begin{equation}
\mathcal{I}_0\dfn(-\infty,s_1],\ \mathcal{I}_1\dfn(s_1,s_2], \mbox{ ... } \mathcal{I}_j\dfn(s_j,s_{j+1}],   \mbox{ ..., } \mathcal{I}_{m_t}\dfn(s_{m_t-1},s_{m_t}],\ \mathcal{I}_{m_t}\dfn(s_{m_t},+\infty)
\end{equation}
and denoting as $L_{i,i+1}(x)$ the straight line passing through the points $(s_{i},V(s_{i}))$ and $(s_{i+1},V(s_{i+1}))$ for $1 \le i \le m_t$, a suitable function $W_{t}(x)$ can be expressed as  
\begin{equation}
\label{ARSderfreeEq}
W_t(x)\dfn
\begin{cases}
	L_{1,2}(x), & x \in \mathcal{I}_0 = (-\infty, s_1];\\
	L_{2,3}(x), & x \in \mathcal{I}_1 = (s_1, s_2];\\
	\min\{L_{j-1,j}(x),L_{j+1,j+2}(x)\}, & x \in \mathcal{I}_j = (s_{j}, s_{j+1}],\ 2 \le j \le m_t-2;\\
	L_{m_t-2,m_t-1}(x), & x \in \mathcal{I}_{m_t-1} = (s_{m_t-1}, s_{m_t}];\\
	L_{m_t-1,m_t}(x), & x \in \mathcal{I}_{m_t} = (s_{m_t}, +\infty ).\\
\end{cases}
\end{equation}
It is apparent also in this case that  $W_t(x)\geq V(x)$, $\forall x\in \mathcal{D}$ and $\forall t\in\mathbb{N}$, as shown in Figure \ref{figARStwoProc}(b).
Therefore, building $W_t(x)$ such that $W_t(x)\geq V(x)$, we have in both cases
\begin{equation}
\label{ARSenvelope}
\pi_t(x) \dfn \exp(W_t(x)) \geq p(x)=\exp\{V(x)\}.
\end{equation}
Since $W_t(x)$ is a piecewise linear function, we obtain an exponential-type proposal density, $\tilde{\pi}_t(x)\propto \pi_t(x)$.
Note also that knowledge of the area below each exponential piece, $k_i=\int_{x\in \mathcal{I}_i} \exp(W_t(x)) dx$ for $0 \le i \le m_t$, is required to generate samples from $\tilde{\pi}_t(x)$, and, given the functional form of $\pi_t(x)$, it is straightforward to compute each of these terms, as well as the normalization constant, $1/c_{\pi}$ with $c_{\pi}=\int_{x\in\mathcal{D}} \pi_t(x)dx$, since $c_{\pi} = k_0 + k_2 + ... + k_n$.

Therefore, we conclude that, since $\pi_t(x)$ is a piecewise-exponential function, it is very easy to draw samples from $\tilde{\pi}_t(x)=\frac{1}{c_{\pi}}\pi_t(x)$ and, since $\pi_t(x) \geq p(x)$, we can apply the rejection sampling principle.
Moreover, when a sample $x'$ drawn from $\tilde{\pi}_t(x)$ is rejected, we can incorporate it into the set of support points, i.e. $\mathcal{S}_{t+1} = \mathcal{S}_t \cup \{x'\}$ and $m_{t+1}=m_t+1$. Then, we can compute a refined approximation, $W_{t+1}(x)$, following one of the two approaches described above, and a new proposal density, $\pi_{t+1}(x)=\exp(W_{t+1}(x))$, can be constructed.
Table \ref{ARS} summarizes the procedure followed by the ARS algorithm to generate $N$ samples from a target $p(x)$.
\begin{table*}[!hbt]
\begin{center}
\caption{Adaptive Rejection Sampling Algorithm.}
\label{ARS}
\begin{tabular}{||l||}
\hline
\hline
1. Start with $i=0$, $t=0$, $m_0=2$, $\mathcal{S}_{0}=\{s_{1}, \ s_{2}\}$ where $s_{1}<s_{2}$ and $[s_1,s_2]$ contains the mode. \\
\ \ \  Let $N$ be the number of desired samples from $p_o(x)\propto p(x)$. \\
2. Build the piecewise-linear function $W_{t}(x)$ as in Eq.~(\ref{ARSwithDer}) or (\ref{ARSderfreeEq}), and as shown in Figure \ref{figARStwoProc}(a) or \ref{figARStwoProc}(b).\\
3. Sample $x'$ from $\tilde{\pi}_{t}(x)\propto \pi_t(x)=\exp(W_{t}(x))$, and $u'$ from $\mathcal{U}([0,1])$. \\
4. If $u'\leq \frac{p(x')}{\pi_{t}(x)}$ then accept $x^{(i)}=x'$ and set $\mathcal{S}_{t+1}=\mathcal{S}_{t}$ ($m_{t+1}=m_t$), $i=i+1$. \\
5. Otherwise, if $u'> \frac{p(x')}{\pi_{t}(x')}$, then reject $x'$, set $\mathcal{S}_{t+1}=\mathcal{S}_{t}\cup \{x'\}$ ($m_{t+1}=m_t+1$). \\ 
6. Sort $\mathcal{S}_{t+1}$ in ascending order and set $t=t+1$. If $i = N$, then stop. Else, go back to step 2.\\
\hline
\hline
\end{tabular}
\end{center}
\end{table*} 



It is important to observe that the adaptive strategy followed by ARS includes new support points only when and where they are needed, i.e. when a sample is rejected, indicating that the discrepancy between the shape of the target and the proposal pdfs is potentially large.
This discrepancy is measured by computing the ratio $p(x)/\pi_t(x)$.
Since $\pi_t(x)$ approaches $p(x)$ when $t \rightarrow \infty$, the ratio $p(x)/\pi_t(x)$ becomes closer and closer to one, and the probability of adding a new support point becomes closer and closer to zero, thus ensuring that the computational cost remains bounded.
More details about the convergence of ARS are provided below.

 \subsection{Convergence}

Note that every time a sample $x'$ drawn from $\pi_t(x)$ is rejected, $x'$ is incorporated as a support point in the new set $\mathcal{S}_{t+1}=\mathcal{S}_{t}\cup \{x'\}$. As a consequence, a refined lower hull $W_{t+1}(x)$ is constructed, yielding a better approximation of the system's potential function, $V(x)$. In this way, $\pi_{t+1}(x) = \exp(W_{t+1}(x))$ becomes ``closer'' to $p(x)$ and the mean acceptance rate can be expected to become higher. 
More precisely, the probability of accepting a sample $x\in \mathcal{D}$ drawn from $\pi_t(x)$ is
\begin{equation}	
a_t(x)\dfn\frac{p(x)}{\pi_t(x)},
\end{equation}
and we define the acceptance rate at the $t$-th iteration of the ARS algorithm, denoted as $\hat{a}_t$, as the expected value of $a_t(x)$ with respect to the normalized pdf $\tilde{\pi}_t(x)$, i.e.,  
\begin{equation}	
\label{MARate}
\hat{a}_t\dfn E[a_t(x)]=\int_{\mathcal{D}}a_t(x)\tilde{\pi}_t(x)dx=\int_{\mathcal{D}}\frac{p(x)}{\pi_t(x)} \tilde{\pi}_t(x)dx=\frac{1}{c_{\pi}}\int_{\mathcal{D}}p(x) dx=\frac{c_p}{c_{\pi}}\leq 1,
\end{equation}
where $1/c_{\pi}$ and $1/c_p$ are the proportionality constants for $\pi_t(x)$ and $p(x)$ respectively, i.e.
\[
	c_{\pi}=\int_{\mathcal{D}} \pi_t(x)dx, 
\]
 and 
	\[
	c_p=\int_{\mathcal{D}} p(x)dx,
\] 
and we remark that $\frac{c_p}{c_{\pi}}\leq 1$ because $\pi_t(x)\geq p(x)$ $\forall x\in \mathcal{D}$.
Hence, from (\ref{MARate}) we conclude that $\hat{a}_t = 1 \Leftrightarrow c_{\pi} = c_p$. Equivalently, defining the distance between two curves $D_{\pi|p}(t)$ as
\begin{equation}	
\label{Intcero}
	D_{\pi|p}(t) \dfn \int_{\mathcal{D}} \left|\pi_t(x)-p(x)\right| dx=\int_{\mathcal{D}} \left|\exp(W_t(x))-\exp(V(x))\right|dx,
\end{equation}
then we have $\hat{a}_t = 1 \Leftrightarrow D_{\pi|p}(t)=0$.
Note that this distance  $D_{\pi|p}(t)$ measures the discrepancy between the proposal $\pi_t(x)$ and the target $p(x)$ pdfs.
In particular, if $D_{\pi|p}(t)$ decreases the acceptance rate $\hat{a}_t=\frac{c_p}{c_{\pi}}$ increases, and, since $\exp(W_t(x)) \geq \exp(V(x))\ \forall x\in \mathcal{D}$, $D_{\pi|p}(t)=0$ if, and only if, $W_t(x)=V(x)$ almost everywhere. Equivalently, $\hat{a}_t=1$ if, and only if, $\pi_t(x)=p(x)$ almost everywhere. 

\section{Adaptive Rejection Metropolis Sampling Algorithm}
\label{Sect_ARMS}
The disadvantage of the classical ARS technique is that it can only draw samples from univariate log-concave densities.
In general, for non-log-concave and multimodal pdfs it is difficult to build a function $W_t(x)$ that satisfies the condition $W_t(x) \geq V(x)$ for all $x\in \mathcal{D}$.
Recognizing this problem, in \citep{Gilks95} an extension of ARS is suggested to deal with pdfs that are not log-concave by appending a Metropolis-Hastings (MH) algorithm step \citep{Hastings70,Metropolis53,Metropolis49}.
The {\it adaptive rejection Metropolis sampling} (ARMS) first performs an RS test, and the discarded samples are used to improve the proposal pdf, as in the standard ARS technique. However, if the sample is accepted in the RS test, then the ARMS adds another statistical control using the MH acceptance rule. This guarantees that the accepted samples are distributed according to the target pdf, $p(x)$, even if $W_t(x) < V(x)$.
Therefore, unlike ARS, the algorithm produces a Markov chain and the resulting samples are correlated.
Finally, we remark that the ARMS technique can be seen as an adaptive generalization of the {\it rejection sampling chain} proposed in \citep{Tierney91,Tierney94}.

The ARMS algorithm is described in detail in Table \ref{ARMS}. The time index $k$ denotes the iteration of the chain whereas $t$ is the index corresponding to evolution of the sequence of proposal pdfs $\{\pi_t(x)\}_{t\in \mathbb{N}}$. 

The key stages are steps 4 and 5.
On the one hand, in step $4$ of Table \ref{ARMS}, when a sample is rejected by the RS test (it is possible if and only if $\pi_t(x')>p(x')$), ARMS adds this new point to the set $\mathcal{S}_t$ and uses it to improve the construction of $W_{t+1}(x)$ and go back to step 2.
On the other hand, when a sample is initially accepted by the RS test (clearly this always happens if $\pi_t(x')<p(x')$), ARMS enters in step 5 and uses the MH acceptance rule to determine whether it is finally accepted or not.
However, notice that ARMS never incorporates this new point to $\mathcal{S}_t$, even if it is finally rejected by the MH step.
Observe also that, if $\pi_t(x)\geq p(x)$ $\forall x\in\mathcal{D}$ and $\forall t\in\mathbb{N}$, then the ARMS becomes the standard ARS scheme, since the proposed sample is always accepted in step 5.

Moreover, since $\pi_t(x)$ depends on the support points, a more rigorous notation would be $\pi_t(x|\mathcal{S}_t)$.
However, since $\mathcal{S}_t$ never contains the current state of the chain{\color{magenta},} then the ARMS can be considered an adaptive independent MH algorithm and, for this reason, the dependence on $\mathcal{S}_t$ is usually omitted from the notation. Indeed, the ARMS can be also seen as a particular case of the auxiliary variables method in \citep{Besag93}.

Finally, we observe that, although in \citep{Gilks95} the ARMS algorithm seems to be tightly linked to a particular mechanism to construct the sequence of the proposal pdfs, in fact these two issues can be studied separately.
Therefore, in the following section we describe the two procedures proposed in literature to construct the function $W_t(x)$ in a suitable way \citep{Gilks95,Meyer08}.

\begin{table*}[!hbt]
\begin{center}
\caption{Adaptive Rejection Metropolis Sampling Algorithm.}
\label{ARMS}
\begin{tabular}{||l||}
\hline
\hline
1.  Start with $k=0$ (iteration of the chain), $t=0$, $\mathcal{S}_{0}=\{s_{1}, \ldots,s_{m_0}\}$, with $s_1 < s_2 < \ldots < s_{m_0}$, \\ 
\ \ \  and choose a value $x_0$. Let $N$ be the required number of iterations of the Markov chain.\\
2. Build a function $W_{t}(x)$ using the set of points $\mathcal{S}_t$ and one of the two procedures described in Section \ref{Sect_Pro_build_W}\\
\ \ \ \citep{Gilks95,Meyer08}.\\
3. Sample $x'$ from $\tilde{\pi}_{t}(x)\propto \pi_t(x)=\exp(W_{t}(x))$, and $u'$ from $\mathcal{U}([0,1])$. \\
4. If $u' > \frac{p(x')}{\pi_t(x')}$, then discard $x'$, set $\mathcal{S}_{t+1}=\mathcal{S}_{t}\cup\{x'\}$, $m_{t+1}=m_t+1$, sort $\mathcal{S}_{t+1}$ in ascending order, \\
\ \ \ update $t=t+1$, and go back to step 2.\\
5. Otherwise, i.e. if $u' \leq \frac{p(x')}{\pi_t(x')}$, set $x_{k+1}=x'$ with probability \\
$\mbox{ }\mbox{ }\mbox{ }\mbox{ }\mbox{ }\mbox{ }\mbox{ }\mbox{ }\mbox{ }\mbox{ }\mbox{ }\mbox{ }\mbox{ }\mbox{ }\mbox{ }\mbox{ }\mbox{ }\mbox{ }\mbox{ }\mbox{ }\mbox{ }$ $\alpha=\min\left[1, \frac{p(x')\min[p(x_k),\pi_t(x_k)]}{p(x_k)\min[p(x'),\pi_t(x')]}\right],$ \\
\ \ \ or set $x_{k+1}=x_k$ with probability $1-\alpha$. Moreover, set $\mathcal{S}_{t+1}=\mathcal{S}_{t}$, $m_{t+1}=m_t$, $t=t+1$ and $k=k+1$.\\
6. If $k=N-1$, then stop. Else, go back to step 2.\\
\hline
\hline
\end{tabular}
\end{center}
\end{table*}



%

%

%

\subsection{Procedure to build $W_t(x)$ proposed in \citep{Gilks95}}
\label{Sect_Pro_build_W}

\begin{figure}[htb]
\centering
\subfigure[]{\includegraphics[width=6cm]{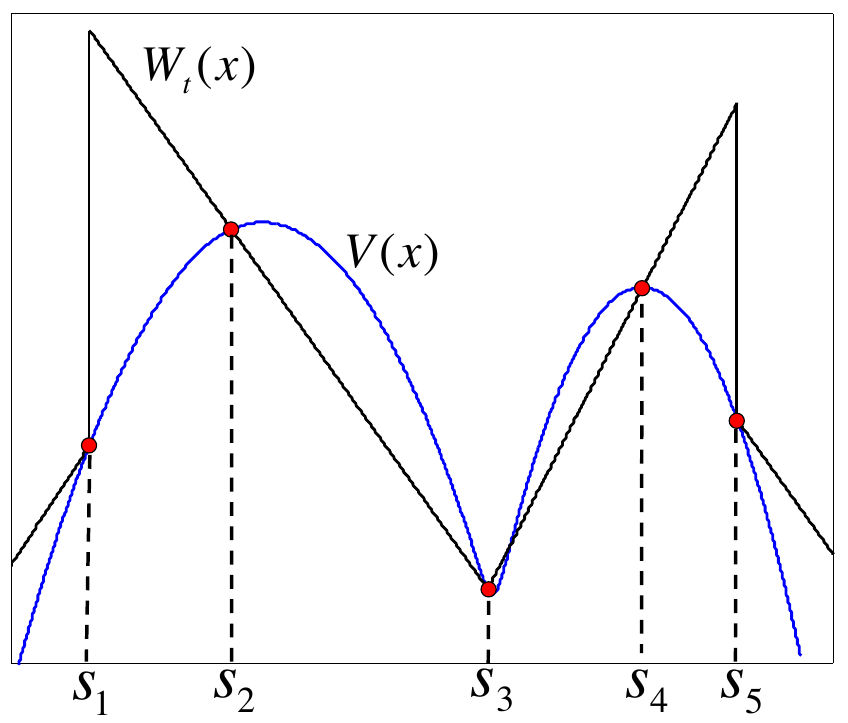}}
 \subfigure[]{ \includegraphics[width=6cm]{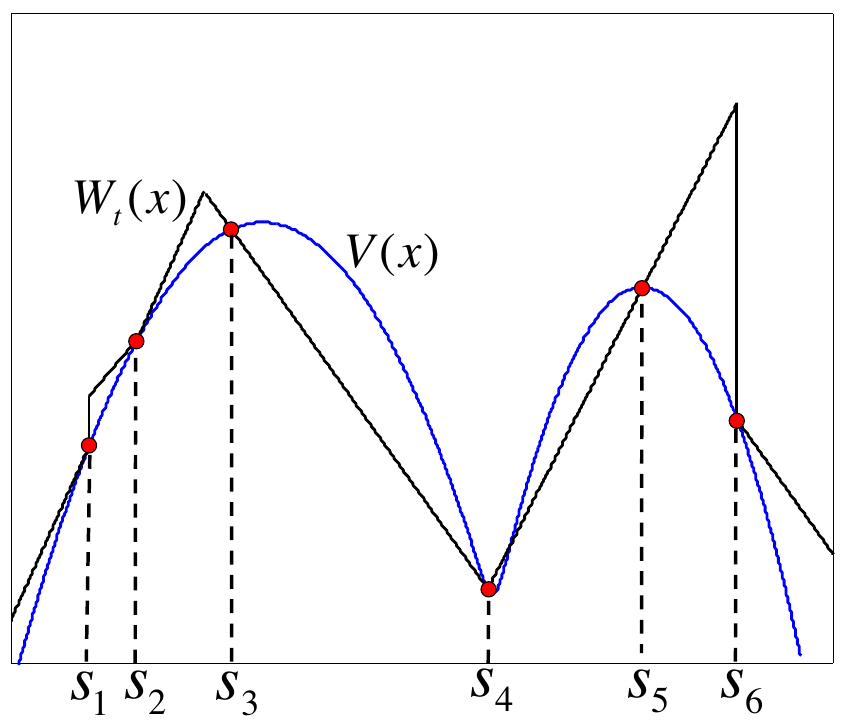}}
  \caption{Example of piecewise linear function $W_t(x)$ built using the procedure described in \citep{Gilks95}: {\bf (a)} with $5$ support points and {\bf (b)} with $6$ support points. Note that $V(x)$ is non-concave and $W_t(x)$ can fall below $V(x)$.}
\label{figARMSProc}
\end{figure}

Consider again a set of support points $\mathcal{S}_t=\{s_1,...,s_{m_t}\}$ and the $m_t+1$ intervals $\mathcal{I}_0=(-\infty,s_1]$, $\mathcal{I}_j=(s_j,s_{j+1}]$, for $j=1,...,m_t-1$ and $\mathcal{I}_{m_t}=(s_{m_t},+\infty)$. Moreover, let us denote as $L_{j,j+1}(x)$ the straight line passing through the points $(s_{j},V(s_{j}))$ and $(s_{j+1},V(s_{j+1}))$ for $j=1,...,m_t-1$, and also set  
$$L_{-1,0}(x)=L_{0,1}(x)\dfn L_{1,2}(x),$$ 
$$L_{m_t,m_t+1}(x)=L_{m_t+1,m_t+2}(x)\dfn L_{m_t-1,m_t}(x).$$  
In the standard ARMS procedure introduced in \citep{Gilks95}, the function $W_t(x)$ is piecewise linear and defined as 
\begin{equation}
\label{EqWtARMS}
W_t(x)=\max\big[L_{i,i+1}(x), \min\left[L_{i-1,i}(x),L_{i+1,i+2}(x)\right] \big]   \mbox{ for }  \mathcal{I}_i=(s_{i}, s_{i+1}],
\end{equation}
with $i=0,\ldots,m_t$. The function $W_t(x)$ can be also rewritten in a more explicit form as 
\begin{equation}
\label{EqWtARMS2}
\small
W_t(x)=
\begin{cases}
L_{1,2}(x), & x \in \mathcal{I}_0=(-\infty, s_1];\\
\max\left\{L_{1,2}(x),L_{2,3}(x)\right\}, & x \in \mathcal{I}_1=(s_{1}, s_{2}];\\
\max\left\{L_{j,j+1}(x), \min\left\{L_{j-1,j}(x),L_{j+1,j+2}(x)\right\} \right\}, & x \in \mathcal{I}_j=(s_{j}, s_{j+1}],
	\ 2 \le j \le m_t-2;\\
\max\left\{L_{m_t-1,m_t}(x),L_{m_t-2,m_t-1}(x)\right\}, & x \in \mathcal{I}_{m_t-1}=(s_{m_t-1}, s_{m_t}];\\
L_{m_t-1,m_t}(x), & x \in \mathcal{I}_{m_t}=(s_{m_t}, +\infty ).
\end{cases}
\end{equation}%
Figure \ref{figARMSProc} illustrates this construction for a generic log-pdf $V(x)$ with (a) $5$ support points and (b) $6$ support points.

It is important to remark that, with this construction, the addition of a new point inside an interval $\mathcal{I}_{i}=(s_i, s_{i+1}]$ also affects the adjacent regions, $\mathcal{I}_{i-1}=(s_{i-1}, s_{i}]$ and $\mathcal{I}_{i+1}=(s_{i+1}, s_{i+2}]$.
Therefore, if the initial proposal pdf $\pi_0(x)$ fulfills the inequality $\pi_0(x)\geq p(x)$ in some subset $\mathcal{C}\subset \mathcal{D}$, the proposal pdf can theoretically be improved in the entire domain $\mathcal{D}$.
However, in practice the convergence of the proposal to the target pdf can be very slow in some cases, and moreover it cannot be ensured in general (see Section \ref{SectStrucLim} for further considerations).

Note also that, if the target pdf $p(x)$ is log-concave, then
\begin{equation*}
	\max\left[L_{j,j+1}(x), \min\left[L_{j-1,j}(x),L_{j+1,j+2}(x)\right] \right] = \min\left[L_{j-1,j}(x),L_{j+1,j+2}(x)\right],
\end{equation*}
for $0 \le j \le m_t$, hence this procedure is reduced to Eq. \eqref{ARSderfreeEq} of the standard ARS technique.
It is important to observe that, although $W_t(x)$ can be expressed in the compact form of Eq.~(\ref{EqWtARMS}), drawing samples from the proposal, $\tilde{\pi}_t(x)\propto \exp(W_t(x))$, requires knowing the equation of each straight line that composes $W_t(x)$, and also the corresponding definition interval (see Figure \ref{figARMSProc}).
Hence, the complexity of drawing from $\tilde{\pi}_t(x)$ is much larger than what equations \eqref{EqWtARMS} or \eqref{EqWtARMS2} might suggest, since it entails the computation of all the intersection points between all the contiguous straight lines (see Figure \ref{figARMSProc}).

Simpler constructions without this requirement could be proposed (see, for instance, Section \ref{SecOtherProc}), but the efficiency of the standard ARMS scheme would be severely compromised, since the good performance of ARMS depends critically on the construction of the proposal, as discussed in the following section and shown in the simulations (see Section \ref{Sect_SIMU}).

Another possible and more complex procedure for building $W_t(x)$ was proposed in \citep{Meyer08}, involving quadratic approximations of the log-pdf $V(x)$ when it is possible, so that the proposal is also formed by truncated Gaussian pdfs.
Indeed, parabolic pieces passing through $3$ support points are used if only if the resulting quadratic function is concave (in order to obtain truncated Gaussian as proposal pdf in the corresponding interval). Otherwise, linear functions are used to build $W_t(x)$. 
The main advantage of this alternative procedure \citep{Meyer08} is that it provides a better approximation of the function $V(x)=\log(p(x))$. However, it does not overcome the critical structural limitation of the standard ARMS technique that we explain in the sequel.


\section{Structural limitation in the adaptive procedure of ARMS}
\label{SectStrucLim}
Although ARMS can be a very effective technique for sampling from univariate non-log-concave pdfs, its performance depends critically on the following two issues:
\begin{itemize}

\item[a)] The construction of $W_t(x)$ should be such that the condition $W_t(x)\geq V(x)$ is satisfied for most intervals $x\in\mathcal{I}_j$ as possible, with $j = 0,\ 1,\ \ldots,\ m_t$.
That is, the proposal function $\pi_t(x)=\exp(-W_t(x))$ must stay above the target pdf $p(x)$, inside as many intervals as possible, covering as much of the domain $\mathcal{D}$ as possible.
In this case the adaptive procedure works almost in the entire domain $\mathcal{D}$ and the proposal density can be improved virtually everywhere.
Indeed, if $\pi_t(x) \geq p(x)$ for all $x\in\mathcal{D}$ and for all $t\in \mathbb{N}$, the ARMS is reduced the classical ARS algorithm.

\item[b)] The addition of a support point within an interval, $\mathcal{I}_j=(s_j,s_{j+1}]$, with $j\in\{0,...,m_t\}$, must entail an improvement of the proposal pdf inside other neighboring intervals when building $W_{t+1}(x)$.
This allows the proposal pdf to improve even inside regions where $\pi_t(x) < p(x)$ and a support point can never be added at the $t$-th iteration, since we could have $\pi_{t+1}(x) > p(x)$ by adding a support point in an adjacent interval.
For instance, in the procedure described in Section \ref{Sect_Pro_build_W} \citep{Gilks95}, when a support point is added inside $\mathcal{I}_j$, the proposal pdf also changes in the intervals $\mathcal{I}_{j-1}$ and $\mathcal{I}_{j+1}$.
Consequently, the drawback of not adding support points within the intervals where $\pi_t(x)< p(x)$ is reduced, but may not completely eliminated, as shown through an example below.
\end{itemize}
In any case, it is important to remark that the convergence of the proposal $\pi_t(x)$ to the target pdf $p(x)$, cannot be guaranteed using any suitable construction of $W_t(x)$ except for the special case where $W_t(x) \ge V(x)$ $\forall x \in \mathcal{D}$ and $\forall t \in \mathbb{N}$, where ARMS is reduced to ARS.
This is due to a fundamental structural limitation of the ARMS technique caused by the impossibility of adding support points inside regions where $\pi_t(x)< p(x)$.

Indeed, it is possible that inside some region $\mathcal{C} \subset \mathcal{D}$ where $\pi_t(x)<p(x)$, we obtain a sequence of proposals $\{\pi_{t+1}(x),\ \pi_{t+2}(x),\ \ldots,\ \pi_{t+\tau}(x)\}$ such that $\pi_{t+1}(x)<p(x),\ \pi_{t+2}(x)<p(x),\ \ldots,\ \pi_{t+\tau}(x)<p(x)$ for an arbitrarily large value of $\tau$, and the discrepancy between the proposal and the target pdfs cannot be reduced after an iteration $t=\tau$, clearly implying that the proposal does not converge to the target for $x \in \mathcal{D}$.

\begin{figure}[htb]
\centering 
\subfigure[]{ \includegraphics[width=6cm]{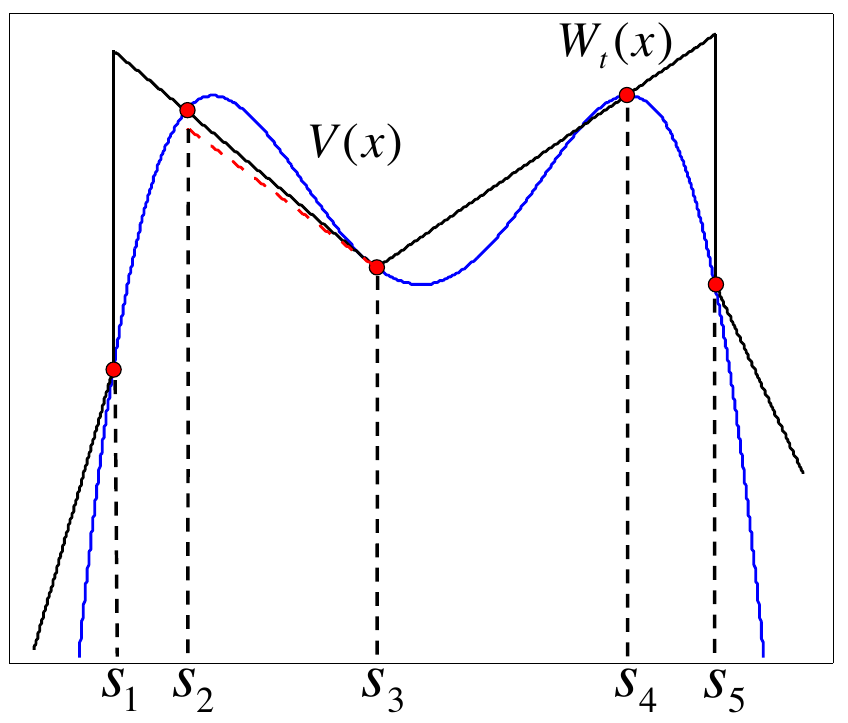}}
 \subfigure[]{ \includegraphics[width=6cm]{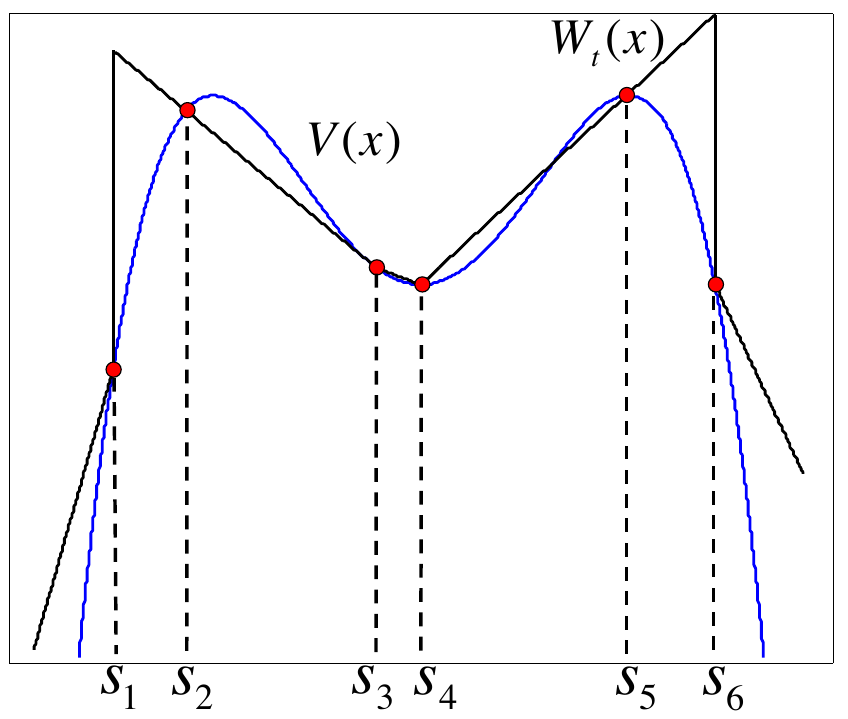}}
 \subfigure[]{ \includegraphics[width=6cm]{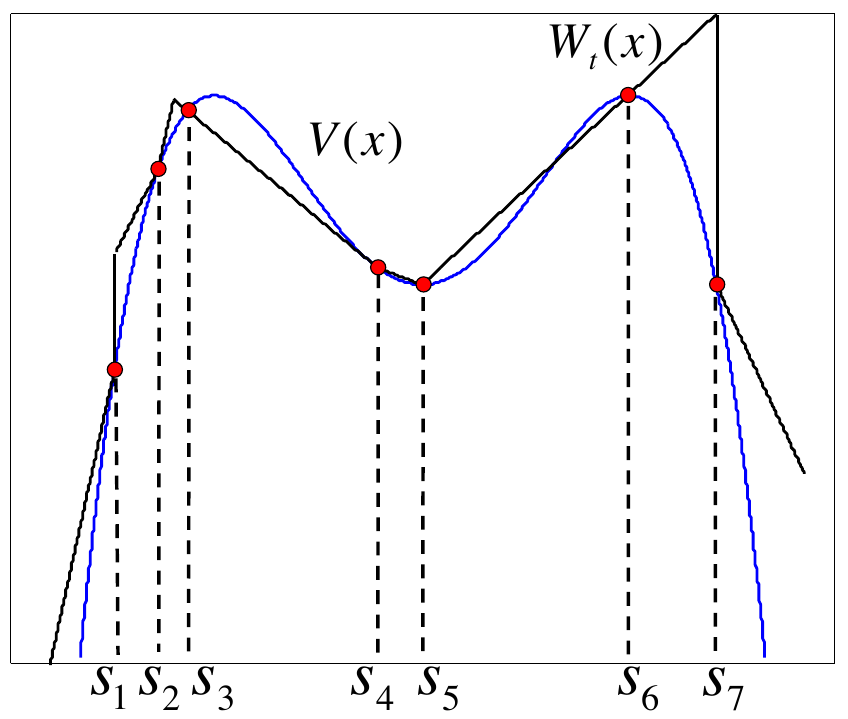}}
 \subfigure[]{ \includegraphics[width=6cm]{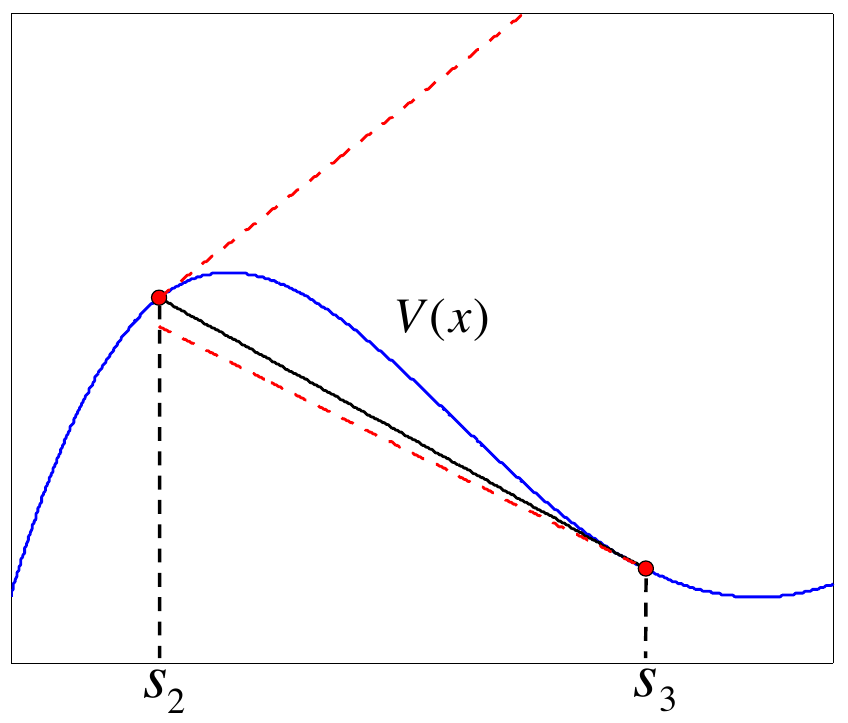}}
  \caption{Example of a critical structural limitation in the adaptive procedure of the ARMS. {\bf{(a)}} Construction of $W_t(x)$ with $5$ support points. Within $\mathcal{I}_2=(s_2,s_3]$ we have $W_t(x)<V(x)$. {\bf(b)}-{\bf (c)} Adding new support points inside the contiguous intervals the construction of $W_t(x)$ does not vary within $\mathcal{I}_2$ ($\mathcal{I}_3$ in Figure (c)). {\bf(d)} The secant line $L_{2,3}(x)$ passing through $(s_2,V(s_2))$ and $(s_3,V(s_3))$, and the two tangent lines to $V(x)$ at $s_2$ and $s_3$.}
\label{figARMSDrawBack}
\end{figure}

In order to illustrate this structural problem of ARMS we provide below an example of an even more critical situation where $\pi_t(x)=\pi_{t+1}(x)=\pi_{t+2}(x)\ldots$ for all $x\in \mathcal{I}_j$, i.e. the proposal pdf does not change within an interval $\mathcal{I}_j\subset \mathcal{D}$.
Consider a multi-modal target density, $p_o(x) \propto p(x) = \exp(V(x))$, with function $V(x)$ as shown in Figure \ref{figARMSDrawBack}(a).
We build $W_t(x)$ using $5$ support points and the procedure of Section \ref{Sect_Pro_build_W} \citep{Gilks95}.
Note that we have $W_t(x)<V(x)$ for all $x$ in the interval $\mathcal{I}_2=(s_2,s_3]$, as shown in Figure \ref{figARMSDrawBack}(a), where the dashed line depicts the tangent line to $V(x)$ at $s_3$.
Figures \ref{figARMSDrawBack}(b) and \ref{figARMSDrawBack}(c) show that we can incorporate new support points ($s_4$ in Figure \ref{figARMSDrawBack}(b) and $s_2$ in Figure \ref{figARMSDrawBack}(c)) arbitrarily close to the interval $\mathcal{I}_3$ ($\mathcal{I}_2$ in Figures \ref{figARMSDrawBack}(a)-(b)) without altering the construction of $W_t(x)$ within $\mathcal{I}_3$ ($\mathcal{I}_2$ in Figures \ref{figARMSDrawBack}(a)-(b)).
The reason is described in the sequel.

Consider Figure \ref{figARMSDrawBack}(a). The construction of $W_t(x)$ for all $x\in\mathcal{I}_2$ depends on the straight lines $L_{1,2}(x)$ passing through $(s_1,V(s_1))$ and $(s_2,V(s_2))$, $L_{2,3}(x)$ passing through $(s_2,V(s_2))$ and $(s_3,V(s_3))$, and $L_{3,4}(x)$ passing through $(s_3,V(s_3))$ and $(s_4,V(s_4))$.
Hence, Eq.~\eqref{EqWtARMS} within this interval can be written as
\begin{equation}
\label{EqWtARMS_drawback}
	W_t(x) = \max\big\{L_{2,3}(x), \min\left\{L_{1,2}(x),L_{3,4}(x)\right\} \big\},
		\qquad \text{for} \quad x \in \mathcal{I}_2=(s_{2}, s_{3}].
\end{equation}
Looking back at Figure \ref{figARMSDrawBack}(a) it becomes apparent that $\min\left\{L_{1,2}(x),L_{3,4}(x)\right\} = L_{3,4}(x)$ and $\max\{L_{2,3}(x), L_{3,4}(x)\} = L_{2,3}(x)$.
Therefore, $W_t(x) = L_{2,3}(x)$ for all $x\in\mathcal{I}_2=[s_2,s_3]$, and this situation does not change when new support points are added inside the contiguous intervals, as shown in Figures \ref{figARMSDrawBack}(c)-(d).
Indeed, consider now the limit case where two points are incorporated arbitrarily close to $s_2$ and $s_3$, so that we can consider the new support points located exactly at $s_2$ and $s_3$.
In this extreme situation the secant lines of the adjacent intervals become tangent lines, as shown in Figure \ref{figARMSDrawBack}(d), and the minimum between the two tangent lines is represented by the straight line tangent to $s_3$.
Moreover, note that this tangent line stays always below the secant line $L_{2,3}(x)$ passing through $(s_2,V(s_2))$ and $(s_3,V(s_3))$, meaning that $W_t(x)=L_{2,3}(x)$ even in this extreme case.

Therefore, with the adaptive procedure used by the standard ARMS technique to build the proposal pdf, the function $W_t(x)$ could remain unchanged in a subset of the entire domain $\mathcal{D}$ that contains a mode of the target pdf, as shown in the previous example.
Hence, the discrepancy between the proposal $\pi_t(x)=\exp(W_t(x))$ and the target $p(x)=\exp(V(x))$, is not reduced at all, as $t\rightarrow +\infty$.

A trivial solution for this drawback of the standard ARMS algorithm could be adding new support points within the set $\mathcal{S}_t$ each time that the MH step in the ARMS (step 5) is applied.
Unfortunately, from a practical point of view the first problem of this approach is the unbounded increase in computational cost since the number of points in $\mathcal{S}_t$ grows indefinitely.
Another major problem from a theoretical point of view is that the convergence of the Markov chain to the invariant pdf cannot be ensured in this case, since the current state of the chain could be contained in $\mathcal{S}_t$ (see \cite[Chapter 8]{Liang10}, \citep{Holden09} for a discussion of this issue).

For these two reasons in the following section we propose two variants of the standard ARMS technique that ensure the convergence of the Markov chain to the target density while keeping the computational cost bounded and low.

\section{Variants of the ARMS algorithm}
\label{Variants_Sect}

In this section we describe two alternative strategies to improve the standard ARMS algorithm.
We denote these two variants as A$^2$RMS and IA$^2$RMS where the A$^2$ emphasizes that we incorporate an additional adaptive step to improve the proposal density w.r.t. the standard ARMS.
Note that in this section we use the more rigorous notation for the proposal, $\pi_t(x|\mathcal{S}_t)$ instead of simply $\pi_t(x)$, for clarity with respect to the convergence of the Markov chain.
 
\subsection{First scheme: A$^2$RMS}

A first possible enhancement of the ARMS technique is summarized below.
The basic underlying idea is enhancing the performance of ARMS by introducing the possibility of adding a sample to the support set even when it is initially accepted by the RS test.
The procedure followed allows A$^2$RMS to incorporate support points inside regions where $\pi_t(x|\mathcal{S}_t) < p(x)$ in a controlled way (i.e. with a decreasing probability as the Markov chain evolves), thus ensuring the convergence to the target and preventing the computational cost from growing indefinitely.

The steps taken by the A$^2$RMS algorithm are the following:
\begin{itemize}

\item[1.] Set $k=0$ (iteration of the chain), $t=0$, choose an initial value $x_0$ and the time to stop the adaptation, $K$. Let $N>K$ be the needed number of iterations of the Markov chain.

\item[2.]  Given a set of support points, $\mathcal{S}_t=\{s_1,\ldots,s_{m_t}\}$ such that $s_1<s_2<\ldots<s_{m_t}$, build an approximation $W_t(x)$ of the potential function $V(x)=\log\left(p(x)\right)$ using a convenient procedure (e.g. the ones described in \citep{Gilks95,Meyer08} or the simpler ones proposed in Section \ref{SecOtherProc}).

\item[3.]  Draw a sample $x'$ from $\tilde{\pi}_t(x|\mathcal{S}_t)\propto \pi_t(x|\mathcal{S}_t)= \exp(W_t)$ and another sample from $u'\sim \mathcal{U}([0,1])$.

\item[4.] If $u' > \frac{p(x')}{\pi_t(x'|\mathcal{S}_t)}$, then discard $x'$, set $\mathcal{S}_{t+1}=\mathcal{S}_{t}\cup\{x'\}$, $m_{t+1}=m_t+1$, sort $\mathcal{S}_{t+1}$ in ascending order, update $t=t+1$, and go back to step 2.

\item[5.] If $u' \leq \frac{p(x')}{\pi_t(x'|\mathcal{S}_t)}$, then: 
\begin{itemize}

\item[5.1.] Set $x_{k+1}=x'$ with probability
\begin{equation}
\label{AlphaEqagora}
	\alpha=\min\left[1, \frac{p(x')\min[p(x_k),\pi_t(x_k|\mathcal{S}_t)]}{p(x_k)\min[p(x'),\pi_t(x'|\mathcal{S}_t)]}\right],
\end{equation}
or $x_{k+1}=x_k$ with probability $1-\alpha$.

\item[5.2.] If $k\leq K$, draw $u_2\sim \mathcal{U}([0,1])$ and if
\begin{equation*}
	u_2> \frac{\pi_t(x'|\mathcal{S}_t)}{p(x')},
\end{equation*}
set $\mathcal{S}_{t+1}=\mathcal{S}_{t}\cup\{x'\}$, $m_{t+1}=m_t+1$ and sort $\mathcal{S}_{t+1}$ in ascending order. Otherwise, set $\mathcal{S}_{t+1}=\mathcal{S}_{t}$, $m_{t+1}=m_t$.
\item[5.3.] Update $t=t+1$ and $k=k+1$.
\end{itemize}
\item[6.] If $k < N$, go back to step 2.
\end{itemize}
We remark again that the key point in the A$^2$RMS algorithm is that, due to the introduction of step 5.2 w.r.t. the ARMS, new support points can also added inside the regions of the domain $\mathcal{D}$ where $\pi_t(x|\mathcal{S}_t)<p(x)$.
Note that, when $\pi_t(x'|\mathcal{S}_t)\geq p(x')$ and we accept the proposed sample $x'$ in the RS test, then we also apply the MH acceptance rule and always accept $x'$, as $\alpha=1$, but we never incorporate $x'$ to the set of support points, since $\frac{\pi_t(x'|\mathcal{S}_t)}{p(x')}\geq 1$ and $u_2\sim \mathcal{U}([0,1])$.
Therefore, if the condition $\pi_t(x|\mathcal{S}_t) \geq p(x)$ is satisfied in all the domain $x\in \mathcal{D}$, then the A$^2$RMS is reduced to the standard ARS, just like it happens for ARMS.

It is important to remark that the effort to code and implement the algorithm is virtually unchanged. Moreover, the ratio $\frac{\pi_t(x'|\mathcal{S}_t)}{p(x')}=\frac{1}{\frac{p(x')}{\pi_t(x'|\mathcal{S}_t)}}$ in the step 5.2 is already calculated in the RS control (steps 4 and 5), hence it is not necessary to evaluate the proposal and the target pdfs again.

Since it is difficult to guarantee that the Markov chain converge to the target pdf, $p_o(x)\propto p(x)$, we have introduced a time $K$ to stop the second adaptation step 5.2. Therefore, theoretically the first $K$ samples produced by the algorithm should be discarded. Observe that for $k>K$ the A$^2$RMS coincides to the standard ARMS.

%
The issue with the convergence of the chain is due to the fact that we are incorporating the current state $x_t$, into the set of support points $\mathcal{S}_t$, on which the proposal depends.
Therefore, the balance condition using the acceptance function in Eq. \eqref{AlphaEqagora} could be not satisfied for $k \le K$ and those first $K$ samples could be seen just as auxiliary random variables obtained as part of the process  to construct a good proposal pdf, and thus should be removed from the set of final returned samples. %
However, it is important to notice the following two facts:

\begin{itemize}

\item[a)] It is a common practice with MCMC techniques to remove a certain amount of the firstly generated samples in order to diminish the effect of the so called \emph{burn-in period}.

\item[b)] We have found out empirically that, even if we set $K=N$ and use all the samples produced by the Markov chain, the A$^2$RMS algorithm also provides very good results, as we show in Section \ref{Sect_SIMU}. Indeed, the probability of incorporating new support points quickly tends to zero as $t$ increases, effectively vanishing as $t \rightarrow +\infty$. This is due to the fact that, each time that a new point is added, the proposal density is improved and becomes closer and closer to the target pdf in the whole domain $\mathcal{D}$. Therefore, the two ratios $\frac{p(x)}{\pi_t(x|\mathcal{S}_t)}$ and $\frac{\pi_t(x|\mathcal{S}_t)}{p(x)}$ approach one and the probability of adding a new support point becomes arbitrarily close to zero.
This implies that the computational cost of the algorithm remains bounded, i.e. a very good approximation of the target can be obtained with a finite number of support points adaptively chosen.  
\end{itemize}
Hence, A$^2$RMS satisfies the first condition needed to ensure the convergence of the Markov chain to the target pdf, known as {\it diminishing adaptation} (see \cite[Chapter 8]{Liang10}, \citep{Haario01,Roberts07}).
Unfortunately, it is difficult to assert whether the A$^2$RMS with $K=N$ also fulfills the second condition needed to guarantee the convergence of the chain, called {\it bounded convergence} \cite[Chapter 8]{Liang10}.
For this reason, although the good results obtained in the numerical simulations with $K=N$ (see Section \ref{Sect_SIMU}) lead us to believe that convergence occurs, it may be safer to set $K<N$ in order to avoid convergence problems in practical applications.

Furthermore, in the next section we  introduce a second A$^2$RMS scheme for which convergence to the target can be ensured, since it is an adaptive {\it independent} MH algorithm \citep{Gasemyr03,Holden09}. 
 
\subsection{Second scheme: Independent A$^2$RMS (IA$^2$RMS)}

A second possible improvement of ARMS is an adaptive independent MH algorithm that we indicate as IA$^2$RMS.
The main modification of IA$^2$RMS w.r.t. A$^2$RMS is building the proposal pdf using possibly all the generated past samples but without taking into account the current state $x_t$ of the chain.
%
%
The IA$^2$RMS is described in the following.
\begin{itemize}

\item[1.] Set $k=0$ (iteration of the chain), $t=0$ and choose an initial value $x_0$.

\item[2.]  Given a set of support points, $\mathcal{S}_t=\{s_1,\ldots,s_{m_t}\}$, such that $s_1<s_2<\ldots<s_{m_t}$, build an approximation $W_t(x)$ of the function $V(x)=\log\left(p(x)\right)$ using a convenient procedure (e.g. the ones described in \citep{Gilks95,Meyer08} or the simpler ones proposed in Section \ref{SecOtherProc}).

\item[3.]  Draw a sample $x'$ from $\tilde{\pi}_t(x|\mathcal{S}_t)\propto \pi_t(x|\mathcal{S}_t)=\exp(W_t)$ and another sample from $u'\sim \mathcal{U}([0,1])$.

\item[4.] If $u' > \frac{p(x')}{\pi_t(x'|\mathcal{S}_t)}$, then discard $x'$, set $\mathcal{S}_{t+1}=\mathcal{S}_{t}\cup\{x'\}$ and $m_{t+1}=m_t+1$, sort $\mathcal{S}_{t+1}$ in ascending order, update $t=t+1$ and go back to step 2. 

\item[5.] If $u' \leq \frac{p(x')}{\pi_t(x'|\mathcal{S}_t)}$, then:
\begin{itemize}

\item[5.1] Set $x_{k+1}=x'$  and $y=x_{k}$ with probability
\begin{equation}
\label{Alfacazzo}
	\alpha=\min\left[1, \frac{p(x')\min[p(x_k),\pi_t(x_k|\mathcal{S}_t)]}{p(x_k)\min[p(x'),\pi_t(x'|\mathcal{S}_t)]}\right],
\end{equation}
or $x_{k+1}=x_k$ and $y=x'$ with probability $1-\alpha$.
\item[5.2] If $y$ is not already contained in $\mathcal{S}_t$, draw $u_2\sim \mathcal{U}([0,1])$, and if
\begin{equation*}
	u_2> \frac{\pi_t(y|\mathcal{S}_t)}{p(y)},
\end{equation*}
set $\mathcal{S}_{t+1}=\mathcal{S}_{t}\cup\{y\}$, $m_{t+1}=m_t+1$ and sort $\mathcal{S}_{t+1}$ in ascending order. Otherwise, set $\mathcal{S}_{t+1}=\mathcal{S}_{t}$ and $m_{t+1}=m_t$.
\item[5.3] Update $t=t+1$ and  $k=k+1$. 
\end{itemize}
\item[6.] If $k<N$ go back to step 2.
\end{itemize}
Note that the main difference of IA$^2$RMS w.r.t. A$^2$RMS lies in steps 5.1 and 5.2, where an auxiliary variable $y$ is introduced to determine whether a new point is added to the set of support points or not instead of using directly $x'$.

This construction leads to a proposal, $\pi_t(x|\mathcal{S}_t)$, which is independent of the current state $x_t$ of the chain, and in \citep{Gasemyr03,Holden09}, \cite[Chapter 8]{Liang10} it is proved that for this kind of adaptive independent MH techniques the convergence of the Markov chain to the invariant density is ensured. 
Once more, if the procedure followed to build the proposal pdf results in $\pi_t(x|\mathcal{S}_t)\geq p(x)$ for all $x\in \mathcal{D}$ and $t \in \mathbb{N}$, then IA$^2$RMS is reduced to the standard ARS algorithm, just like A$^2$RMS and ARMS. 

Moreover, note that also in the  IA$^2$RMS the two components $\pi_t(y|\mathcal{S}_t)$ and $p(y)$, in the ratio $\frac{\pi_t(y|\mathcal{S}_t)}{p(y)}$, are already calculated either if $y=x'$ or $y=x_t$ for instance in Eq. (\ref{Alfacazzo}), hence it is not necessary to evaluate the proposal and the target densities again.


Finally, note that new statistical information is used in the two improved ARMS algorithms proposed (A$^2$RMS and IA$^2$RMS) to update the proposal density also in the regions of the space where $\pi_t(x|\mathcal{S}_t)<p(x)$.
Consequently, these two algorithms are no longer subject to the restrictions of ARMS for attaining a good performance, i.e. that $W_t(x) \ge V(x)$ in most intervals and the improvement of the proposal in adjacent regions when a new support point is added.
Therefore, as shown in the following section, simpler procedures can be used to build the function $W_t(x)$, reducing the overall computational cost of the resulting algorithms and making it easier to code them.

 \section{Alternative procedures to build  $\pi_t(x)$}
\label{SecOtherProc}
As discussed before, the improvements proposed in the structure of the standard ARMS allow us to use simpler procedures to construct the function $W_t(x)$.

\begin{figure}[htb]
\centering 
\subfigure[]{ \includegraphics[width=5cm]{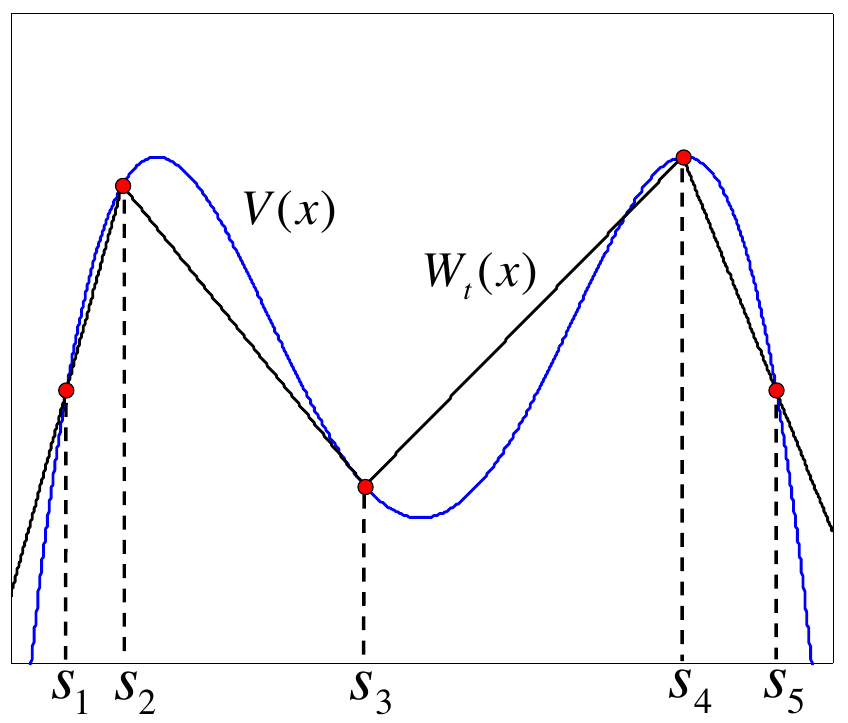}}
 \subfigure[]{ \includegraphics[width=5cm]{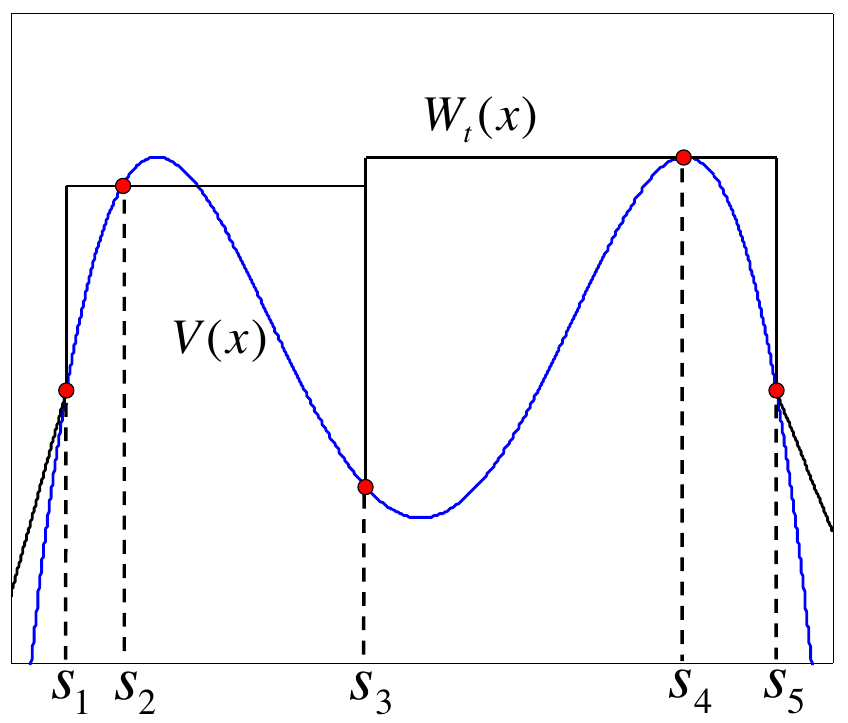}}
 \subfigure[]{ \includegraphics[width=5cm]{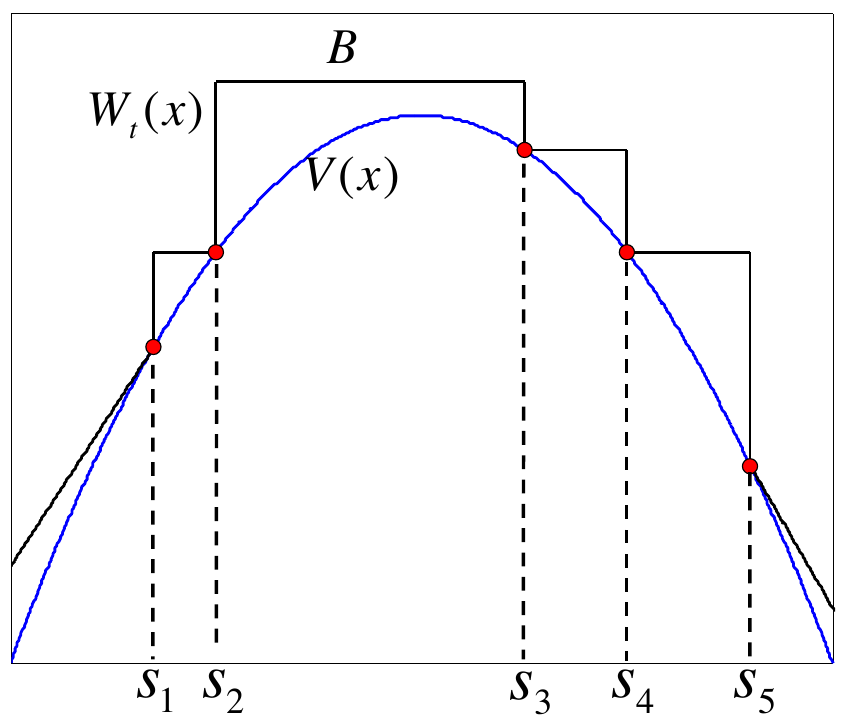}} 
  \caption{{\bf(a)} Example of the construction of $W_t(x)$ using the procedure described in Eq.~(\ref{EqSimpler1}). {\bf(b)} Example of the construction of $W_t(x)$ using the procedure given by Eq.~(\ref{EqSimpler2}). {\bf(c)} If the function $V(x)$ is concave, the procedure in Eq.~(\ref{EqSimpler2}) can be modified to obtain $\pi_t (x)\geq p(x)$ for all $x\in \mathcal{D}$. Indeed, this figure provides an example of the construction of $W_t(x)$ when an upper bound $B \ge V(x)$ is available and the location of the maximum can be estimated.}
\label{figAlternativeProc}
\end{figure}

For instance, a first approach inspired by the one used in ARMS is defining $W_t(x)$ inside the $i$-th interval as the straight line going through $(s_i,V(s_i))$ and $(s_{i+1},V(s_{i+1}))$, $L_{i,i+1}(x)$, for $1 \le i \le m_t-1$, and extending the straight lines corresponding to $\mathcal{I}_1$ and $\mathcal{I}_{m_t-1}$ towards minus and plus infinity for the first and last intervals, $\mathcal{I}_0$ and $\mathcal{I}_{m_t}$ respectively.
Mathematically, this can be expressed as
\begin{equation}
\label{EqSimpler1}
	W_t(x)=
		\begin{cases}
			L_{1,2}(x), & x \in \mathcal{I}_0=(-\infty, s_1];\\
			L_{i,i+1}(x), & x \in \mathcal{I}_i=(s_{i}, s_{i+1}],\ 1 \le i \le m_t-1;\\
			L_{m_t-1,m_t}(x), & x \in \mathcal{I}_{m_t}=(s_{m_t},+\infty).
		\end{cases}
\end{equation}
This is illustrated in Figure \ref{figAlternativeProc}(a).
Note that, although this procedure looks similar to the one used in ARMS, as described by Eqs. \eqref{EqWtARMS} and \eqref{EqWtARMS2}, it is much simpler in fact, since there is not any minimization or maximization involved, and thus it does not require the calculation of intersection points to determine when one straight line is above the other.

Moreover, an even simpler procedure to construct $W_t(x)$ can be devised from \eqref{EqSimpler1}: using a piecewise constant approximation with two straight lines inside the first and last intervals. Mathematically, it can be expressed as
\begin{equation}
\label{EqSimpler2}
	W_t(x)=
		\begin{cases}
			L_{1,2}(x), & x \in \mathcal{I}_0=(-\infty, s_1];\\
			\max\left\{V(s_{i}),V(s_{i+1}) \right\}, & x \in \mathcal{I}_i=(s_{i}, s_{i+1}];\\
			L_{m_t-1,m_t}(x), & x \in \mathcal{I}_{m_t}=(s_{m_t}, +\infty ).
		\end{cases}
\end{equation}
The construction described above leads to the simplest proposal density possible, i.e., a collection of uniform pdfs with two exponential tails.
Figure \ref{figAlternativeProc}(b) shows an example of the construction of the proposal using this approach.

Note that, even when the target pdf is log-concave, using these two procedures the resulting ARMS-type techniques are not reduced to standard ARS, since there is no guarantee that the resulting proposal is above the target pdf.
Indeed, when $V(x)=\log[p(x)]$ is concave, the first procedure described by Eq.~\eqref{EqSimpler2} will produce a proposal which is always below the target except inside the first and last intervals.
However, the second procedure described by Eq.~\eqref{EqSimpler2} can be easily modified to yield $\pi_t(x) = \exp(W_t(x)) \geq p(x) = \exp(V(x))\ \forall x \in \mathcal{D}$ for log-concave target densities, if the position of the mode is known or if an upper bound for $V(x)$ is available and the location of the mode can be estimated (e.g. using the sign of the first derivative).
Figure \ref{figAlternativeProc}(c) depicts an example of the construction of $W_t(x)$ when $V(x)$ is concave using an upper bound $B \geq V(x)$ to obtain $\pi_t (x)\geq p(x)\ \forall x\in \mathcal{D}$.
Moreover, it is important to remark that the procedure in Eq.~\eqref{EqSimpler2}, without any modifications, converges virtually to produce $\pi_t (x) = \exp(W_t(x)) \geq p(x) = \exp(V(x))$ almost everywhere when $V(x)$ is concave and $t \rightarrow +\infty$. 

Finally, we note that we could also apply the procedures proposed for {\it adaptive triangular Metropolis sampling (ATRIMS)} and {\it adaptive trapezoid Metropolis sampling} (ATRAMS) to build the proposal, even though the structure of these two algorithms is completely different to ARMS \citep{Cai08}.
In both cases the proposal is constructed {\it directly} in the domain of the target pdf, $p(x)$, rather than in the domain of the log-pdf, $V(x)=\log(p(x))$.

For instance, the basic idea proposed for ATRAMS is using straight lines, $\widetilde{L}_{i,i+1}(x),$\footnote{Note that we use the symbol $\sim$ to distinguish $\widetilde{L}_{i,i+1}(x)$ and $L_{i,i+1}(x)$. Indeed, $\widetilde{L}_{i,i+1}(x)$ is a straight line built directly in the domain of $p(x)$ whereas the linear function $L_{i,i+1}(x)$ is constructed in the log-domain.} passing through the points $(s_{i},p(s_{i}))$ and $(s_{i+1},p(s_{i+1}))$ for $i=1,\ldots,m_t-1$ and two exponential pieces, $E_0(x)$ and $E_{m_t}(x)$, for the tails:
\begin{equation}
\label{EqSimpler3}
	\tilde{\pi}_t(x)\propto \pi_t(x)=
		\begin{cases}
			E_{0}(x), & x \in \mathcal{I}_0 =(-\infty, s_1];\\
			\widetilde{L}_{i,i+1}(x), & x \in \mathcal{I}_i = (s_{i},s_{i+1}],\ i=1,\ldots,m_t-1;\\
			E_{m_t}(x), & x \in \mathcal{I}_{m_t} = (s_{m_t},+\infty).\\
\end{cases}
\end{equation}
Unlike in \citep{Cai08}, here the tails $E_0(x)$ and $E_{m_t}(x)$ do not necessarily have to be equivalent in the areas they enclose.
Indeed, we may follow a much simpler approach calculating two secant lines $L_{1,2}(x)$ and $L_{m_t-1,m_t}(x)$  passing through $(s_1,V(s_1))$, $(s_{2},V(s_{2}))$, and $(s_{m_t-1},V(s_{m_t-1}))$, $(s_{m_t},V(s_{m_t}))$ respectively, so that the two exponential tails are defined as $E_0(x)=\exp\{L_{1,2}(x)\}$ and $E_{m_t}(x)=\exp\{L_{m_t-1,m_t}(x)\}$.

\begin{figure}[htb]
\centering 
 \subfigure[]{\includegraphics[width=5cm]{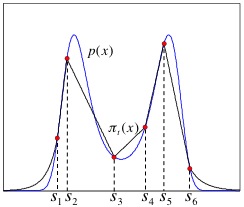}}
\subfigure[]{\includegraphics[width=5cm]{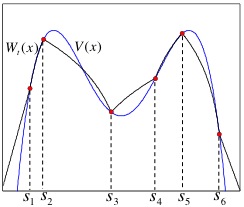}}
  \caption{{\bf(a)} Example of the construction of the proposal pdf, $\tilde{\pi}_t(x) \propto \pi_t(x)$, using a procedure described in \citep{Cai08} within the ATRAMS algorithm. {\bf (b)} The same construction in the log-domain, i.e. for $V(x)=\log(p(x))$ and $W_{t}(x)=\log(\pi_t(x))$.}
\label{figAlternativeProc2}
\end{figure}

Figure \ref{figAlternativeProc2} depicts an example of the construction of $\pi_t(x)$ using this last procedure.
Note that drawing samples from these trapezoidal pdfs inside $\mathcal{I}_i=(s_i,s_{i+1}]$ can be easily done \citep{Cai08,Devroye86}.
Indeed, given $u',v'\sim\mathcal{U}([s_{i},s_{i+1}])$ and $w' \sim \mathcal{U}([0,1])$, then 
\begin{equation}
	x'=
		\begin{cases}
			\min\{u',v'\}, &  w' < \frac{p(s_{i})}{p(s_{i})+p(s_{i+1})};\\
			\max\{u',v'\}, & w' \ge \frac{p(s_{i})}{p(s_{i})+p(s_{i+1})};\\
		\end{cases}
\end{equation}
is distributed according to a trapezoidal density defined in the interval $\mathcal{I}_i=[s_i,s_{i+1}]$.

All these procedures reduce the computational cost of the so called ARMS-type algorithms. However, it is important to remark that they could be inefficient within the standard ARMS structure without the proposed enhancements,  as shown in Section \ref{Sect_SIMU} and due to the reasons explained in Section \ref{Sect_ARMS}.
An important exception is the simple procedure described by Eq.~\eqref{EqSimpler2}, which leads to very good results also in combination with the standard ARMS (although even better results are obtained for A$^2$RMS and IA$^2$RMS), probably due to the fact that the proposal stays above the target almost everywhere.

\subsection{Initial support points and tails}
Depending on the type of construction, the minimum possible number $m_0$ of initial support points is at least $2$ with the procedures described above or $3$ with the procedure in \citep{Gilks95}. In general, to avoid numerical problems the user must only assure that $p(s')>0$ for at least one support point $s'$. However, to speed up the convergence and render the algorithm more robust (avoiding any kind of numerical problems), a greater number of support points can be used. Moreover, observe that there is clearly a trade-off between the initial performance and overall computational cost.

It is important to note that the dependence on the choice of the set $\mathcal{S}_0$ can be reduced adding more probability in the tails of the proposal PDF, initially. Indeed, consider the two slopes  $l_{t,1}$ and $l_{t,2}$ of the linear functions that form $W_t(x)$ in $\mathcal{I}_0$ and $\mathcal{I}_{m_t+1}$, constructed as explained above\footnote{Note that, in general, a control on the slopes $l_{t,1}$ and $l_{t,2}$ is needed to attain a proper proposal PDF since it is necessary that $l_{t,1}>0$ and $l_{t,2}<0$.}. For instance, we can add more probability in the tails of the proposal just by setting
\begin{gather}
\begin{split}
&l_{t,1}'=l_{t,1}(1-\beta e^{-\alpha t}), \\
&l_{t,2}'=l_{t,2}(1-\beta e^{-\alpha t}), \\
\end{split} 
\end{gather}
where $0\leq\beta \leq 1$, $\alpha>0$, and $l_{t,i}'\rightarrow l_{ti}$, $i=1,2$, for $t\rightarrow +\infty$.
Moreover, the proposed algorithms work even if the target PDF has heavy tails on account of the second test in A$^2$RMS and IA$^2$RMS that allows the incorporation of new support points where $p(x)> \pi_t(x|\mathcal{S}_t)$.
 However, if we know what kind of tails the target distribution has, we can speed up the convergence and reduce the burn-in period by changing the construction of the tails of the proposal. For instance, we could use a function of type $\log[1/x^{\gamma}]$, with $\gamma>1$,  to build $W_t(x)$ in $\mathcal{I}_0$ and $\mathcal{I}_{m_t+1}$, instead of straight lines.

 
\section{Simulations}
\label{Sect_SIMU}

In this  section we compare the performance of the two proposed algorithms with the standard ARMS scheme using four alternative procedures to build the proposal pdf $\pi_t(x)$ in all cases: the one proposed by standard ARMS and described by Eq. \eqref{EqWtARMS} (procedure 1), the two simpler procedures given by Eqs. \eqref{EqSimpler1} and \eqref{EqSimpler2} (procedures 2 and 3, respectively), and the procedure described in Eq. \eqref{EqSimpler3}, which is adapted from the ATRAM technique \citep{Cai08} (procedure 4).\footnote{Matlab code using the procedure in Eq. \eqref{EqSimpler2} to build the proposal is provided at \texttt{http://a2rms.sourceforge.net/}.} 

For the simulations, we consider a target pdf, $p_o(x) \propto p(x)$, generated as a mixture of $3$ Gaussian densities,
\begin{equation}
p_o(x)= 0.3 \mathcal{N}(x;-5,1)+0.3 \mathcal{N}(x;1,1)+0.4 \mathcal{N}(x;7,1), 
\end{equation}
where $N(x;\mu, \sigma^2)$ denotes a Gaussian pdf with mean $\mu$ and variance $\sigma^2$. 
%
In order to analyze the accuracy of the different algorithms, we estimate the mean of $p_o(x)$ using the generated samples, comparing it to the true value, $E\{p_o(x)\} = 1.6$.
Moreover, we also provide an estimation of the linear correlation among consecutive samples of the Markov chain, an approximation of the distance $D_{\pi|p}(t)$ in Eq. \eqref{Intcero} between the constructed proposal and the target pdf, the number of rejections and the number of support points at each iteration.

We consider $N=5000$ iterations of the Markov chain and use an initial set $\mathcal{S}_0=\{s_{1}=-10,s_2=a,s_3=b, s_4=10\}$ formed by $m_0=4$ support points, where $a$ and $b$ (with $a<b$) are chosen randomly inside the interval $[-10,10]$ in all cases.\footnote{Note that, in general, it is necessary to check that the slope of the proposal is positive inside $\mathcal{I}_0$ and negative inside $\mathcal{I}_{m_t}$ in order to obtain a proper proposal pdf. Therefore, we fix $s_1 = -10$ and $s_4 = 10$ just to guarantee this for $t=0$, thus avoiding the problem of having to control the slopes of $W_t(x)$ inside the first and last intervals, $\mathcal{I}_0=(-\infty,s_1]$ and $\mathcal{I}_{m_t}=[s_{m_t},+\infty)$ respectively. Note also that support points smaller than -10 or larger than +10 can be added as part of the adaptive construction of the proposal without compromising the slope of the tails in any of the approaches.}
Moreover, in the first proposed scheme (A$^2$RMS) we set $K=N$, i.e. we never stop the adaptation and use all the samples generated by the chain for the estimation.

Tables {\ref{resultsARMS}, \ref{resultsA2RMS} and \ref{resultsIA2RMS}} show the results obtained with the different algorithms (standard ARMS, A$^2$RMS and IA$^2$RMS), averaged over $2000$ runs, using all the $N=5000$ samples generated by the Markov chain and different procedures to build the proposal pdf.  
The columns of the Tables show (from left to right): the procedure used to build the proposal pdf, $\pi_t(x)$; the estimated mean (the true value is $1.6$); the standard deviation of the estimation; the estimated linear correlation between consecutive samples of the chain; the average final number of support points;\footnote{The number of support points is equal to the sum of the number of rejections in the two controls plus the $4$ initial points.} the average number of rejections in the first control (RS test); the average number of rejections in the second control (MH step);\footnote{This test does not exist in the standard ARMS structure. Hence, the value of this column in Table \ref{resultsARMS} is always $0$.} and an approximation of the discrepancy between the target and the proposal pdfs, given by $D_{\pi|p}(t)$, after $N=5000$ iterations.
\begin{table*}[!hbt]
\def\marginwidth{1.5mm}
\begin{center}
\caption{Results with $N=5000$ for the standard ARMS structure}
\label{resultsARMS}
\begin{tabular}{|c@{\hspace{\marginwidth}}|c@{\hspace{\marginwidth}}|c@{\hspace{\marginwidth}}|c@{\hspace{\marginwidth}}|c@{\hspace{\marginwidth}}|c@{\hspace{\marginwidth}}|c@{\hspace{\marginwidth}}|c@{\hspace{\marginwidth}}|}
\hline
{\bf Procedure}  &{\bf Estimated} &{\bf Std of} & {\bf Estimated} &{\bf Avg. num.}&{\bf Avg. num. of} &{\bf Avg. num. of} &{\bf Approx} \\
{\bf to build}  &{\bf mean} &{\bf the}  &{\bf correlation} &{\bf of support} &{\bf rejections in} &{\bf rejections in the} & {\bf of} \\
{\bf$\pi_t(x)$}& & {\bf estimation}  &  & {\bf points}&{\bf the RS test} &{\bf second control} &{\bf $D_{\pi|p}(t)$} \\ 
\hline
\hline
1  &  1.6480 & 0.7301 & {\bf 0.3856} & 65.8717 & 61.8717 & 0 &  {\bf 3.0020}\\
\hline
2  & 1.7541 & 1.0908 & 0.7722 & {\bf 12.0492} & 8.0492 & 0 & 8.0516 \\
\hline
3  &  {\bf 1.5935} & {\bf 0.2300} & 0.6132 & 164.1881 & 160.1881 & 0 & 6.1518 \\
\hline
4  & 1.5670 & 0.4961 & 0.7083 & 37.8231 & 33.8231 & 0 & 7.1339 \\
\hline
\end{tabular}
\end{center}
\end{table*} 


\begin{table*}[!hbt]
\def\marginwidth{1.5mm}
\begin{center}
\caption{Results with $N=5000$ for the A$^{2}$RMS structure}
\label{resultsA2RMS}
\begin{tabular}{|c@{\hspace{\marginwidth}}|c@{\hspace{\marginwidth}}|c@{\hspace{\marginwidth}}|c@{\hspace{\marginwidth}}|c@{\hspace{\marginwidth}}|c@{\hspace{\marginwidth}}|c@{\hspace{\marginwidth}}|c@{\hspace{\marginwidth}}|}
\hline
{\bf Procedure}  &{\bf Estimated} &{\bf Std of} & {\bf Estimated} &{\bf Avg. num.}&{\bf Avg. num. of} &{\bf Avg. num. of} &{\bf Approx} \\
{\bf to build}  &{\bf mean} &{\bf the}  &{\bf correlation} &{\bf of support} &{\bf rejections in} &{\bf rejections in the} & {\bf of} \\
{\bf$\pi_t(x)$}& & {\bf estimation}  &  & {\bf points}&{\bf the RS test} &{\bf second control} &{\bf $D_{\pi|p}(t)$} \\ 
\hline
\hline
1  & 1.6244 &   0.1184 & 0.0038 & 94.4969 & 80 & 10.4969 & 0.0613 \\
\hline
2  & 1.7381 & 0.2258 & 0.0229 &  {\bf 86.1017} & 9.8948 & 72.2069 & {\bf 0.0580} \\
\hline
3  & {\bf 1.5984} & {\bf 0.0797} & {\bf 0.0026} &  317.4166 & 305.0933 & 8.3232 & 0.3110 \\
\hline
4  & 1.6051 & 0.0854 &   0.0060 & 92.3517 & 54.8663 & 33.4855 &  0.0590 \\
\hline
\end{tabular}
\end{center}
\end{table*}
\begin{table*}[!hbt]
\def\marginwidth{1.5mm}
\begin{center}
\caption{Results with $N=5000$ for the IA$^{2}$RMS structure}
\label{resultsIA2RMS}
\begin{tabular}{|c@{\hspace{\marginwidth}}|c@{\hspace{\marginwidth}}|c@{\hspace{\marginwidth}}|c@{\hspace{\marginwidth}}|c@{\hspace{\marginwidth}}|c@{\hspace{\marginwidth}}|c@{\hspace{\marginwidth}}|c@{\hspace{\marginwidth}}|}
\hline
{\bf Procedure}  &{\bf Estimated} &{\bf Std of} & {\bf Estimated} &{\bf Avg. num.}&{\bf Avg. num. of} &{\bf Avg. num. of} &{\bf Approx} \\
{\bf to build}  &{\bf mean} &{\bf the}  &{\bf correlation} &{\bf of support} &{\bf rejections in} &{\bf rejections in the} & {\bf of} \\
{\bf$\pi_t(x)$}& & {\bf estimation}  &  & {\bf points}&{\bf the RS test} &{\bf second control} &{\bf $D_{\pi|p}(t)$} \\ 
\hline
\hline
1  & 1.6233 & 0.1238 &  0.0041 & 94.8435 & 81.1630 & 9.6805 & 0.0609 \\
\hline
2  & 1.7244 & 0.2194 & 0.0203 & {\bf 85.6441} &  10.0169 & 71.6271 & {\bf 0.0565} \\
\hline
3  & {\bf 1.6007}  & {\bf 0.0950} & {\bf 0.0021} & 317.5360 & 306.2528 & 7.2832 & 0.3009 \\
\hline
4  & 1.6011 & 0.1308 &  0.0054 & 92.1333 & 55.0159 & 33.1175 & 0.0582 \\
\hline
\end{tabular}
\end{center}
\end{table*}

\begin{figure}[!H]
\centering 
\centerline{
\subfigure[]{\includegraphics[width=0.3\textwidth]{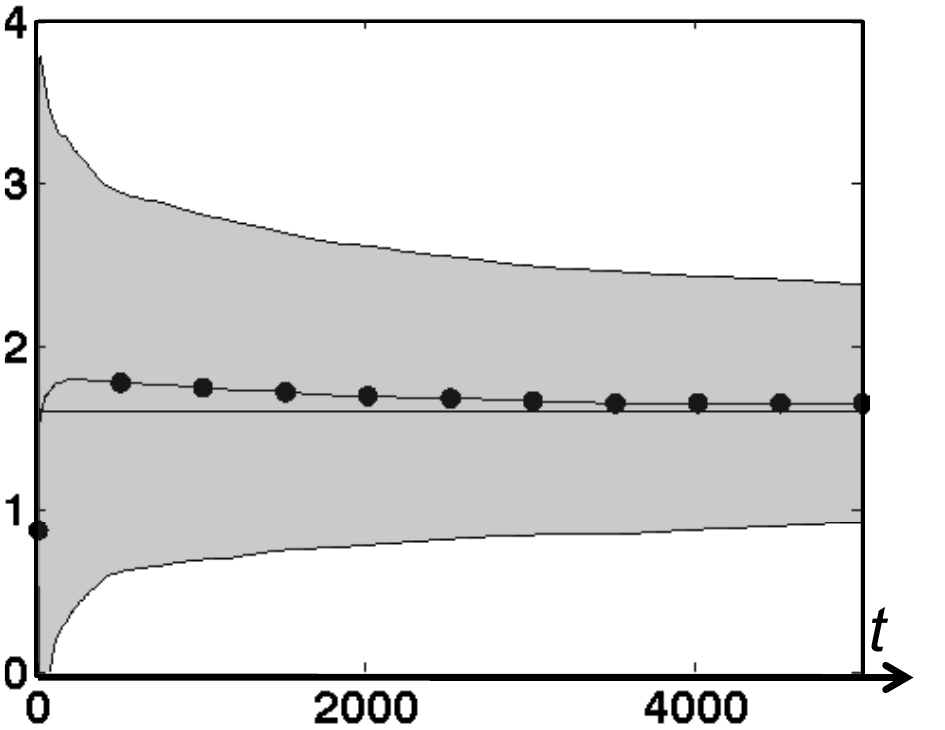}}
 \subfigure[]{ \includegraphics[width=0.3\textwidth]{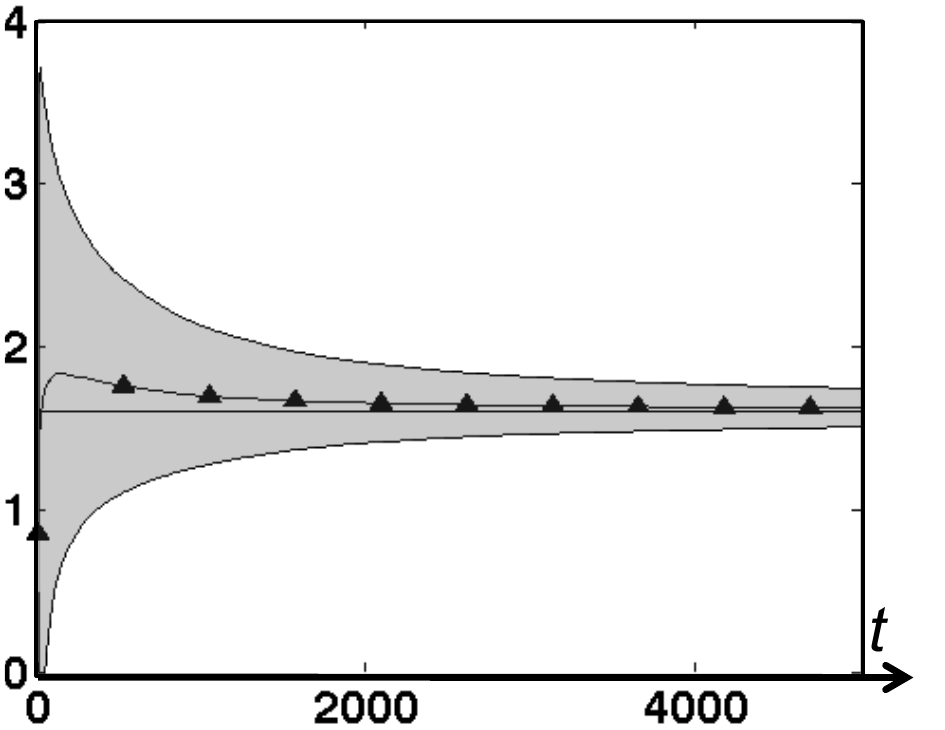}}
 \subfigure[]{\includegraphics[width=0.3\textwidth]{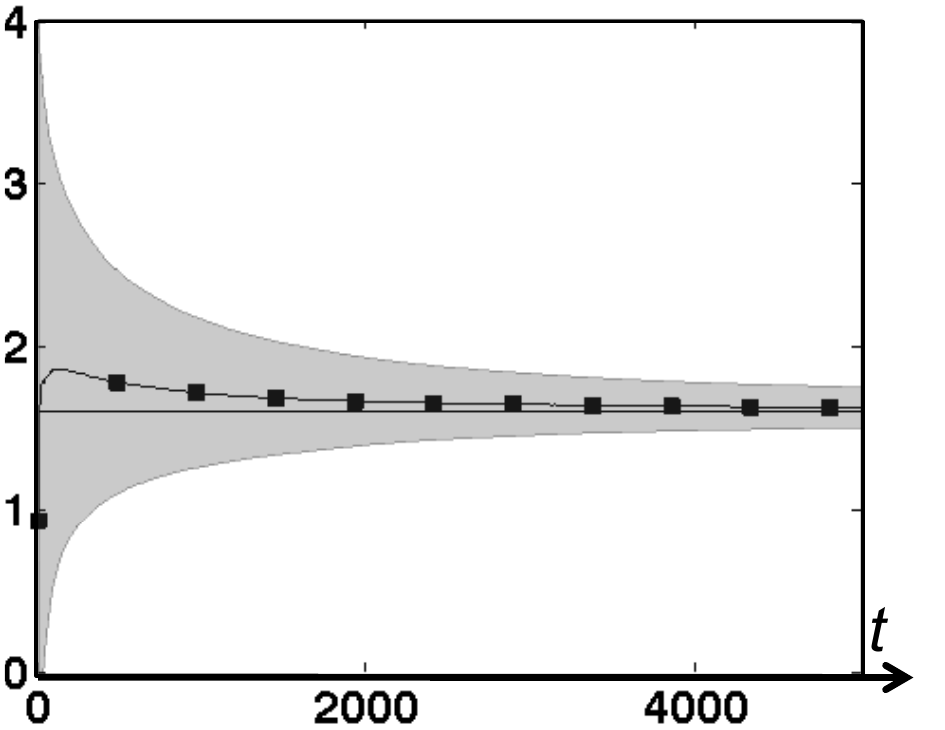}}
}
\centerline{
\subfigure[]{\includegraphics[width=0.3\textwidth]{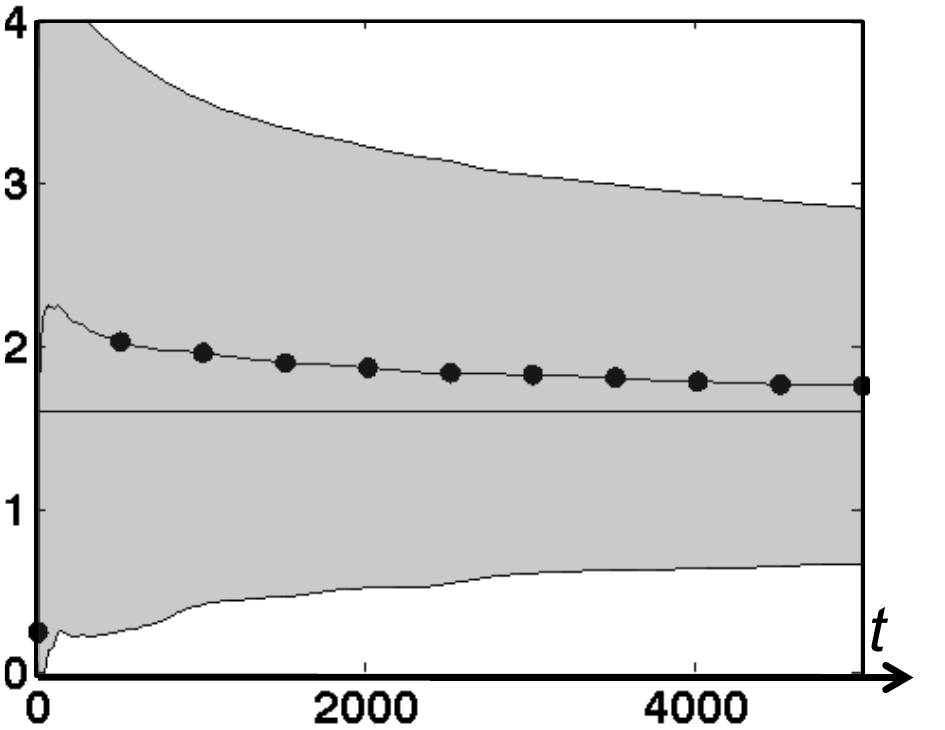}}
 \subfigure[]{ \includegraphics[width=0.3\textwidth]{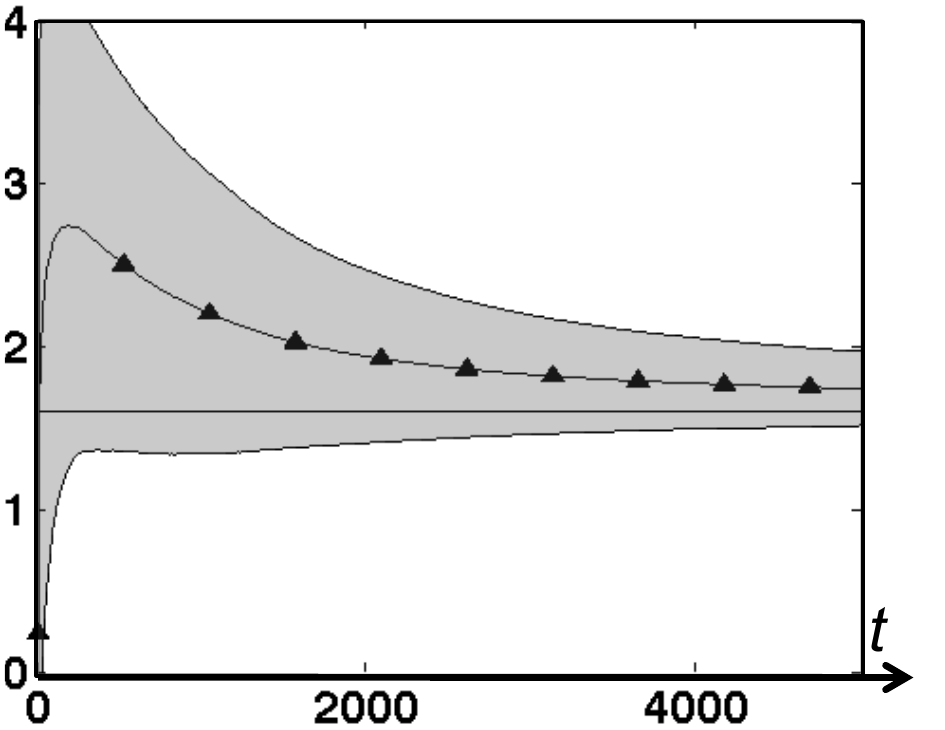}}
 \subfigure[]{\includegraphics[width=0.3\textwidth]{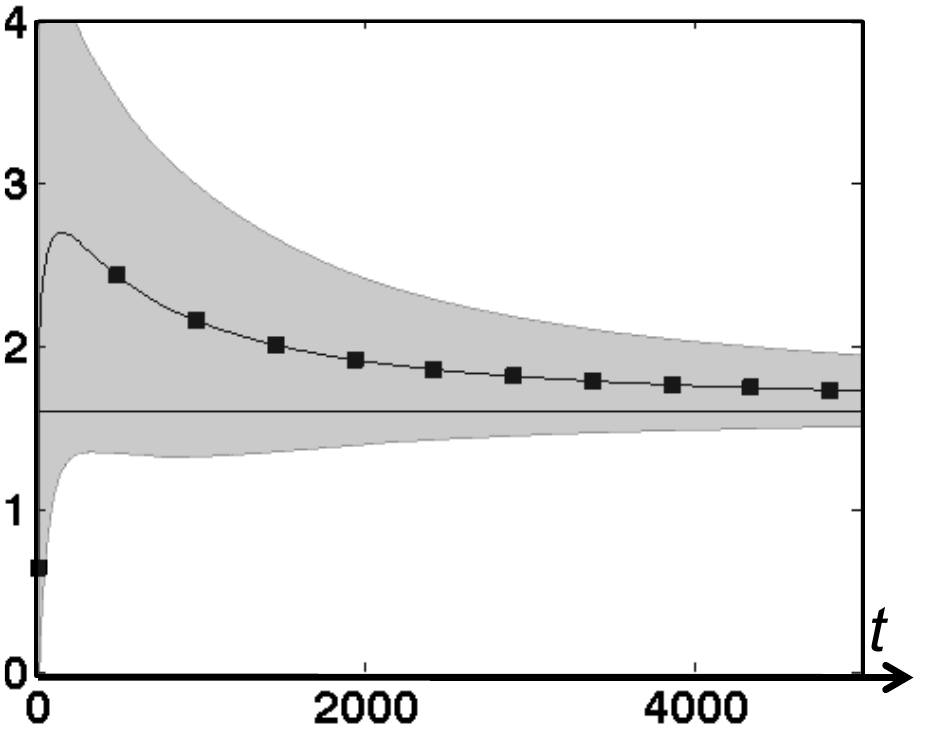}}
}
\centerline{
\subfigure[]{\includegraphics[width=0.3\textwidth]{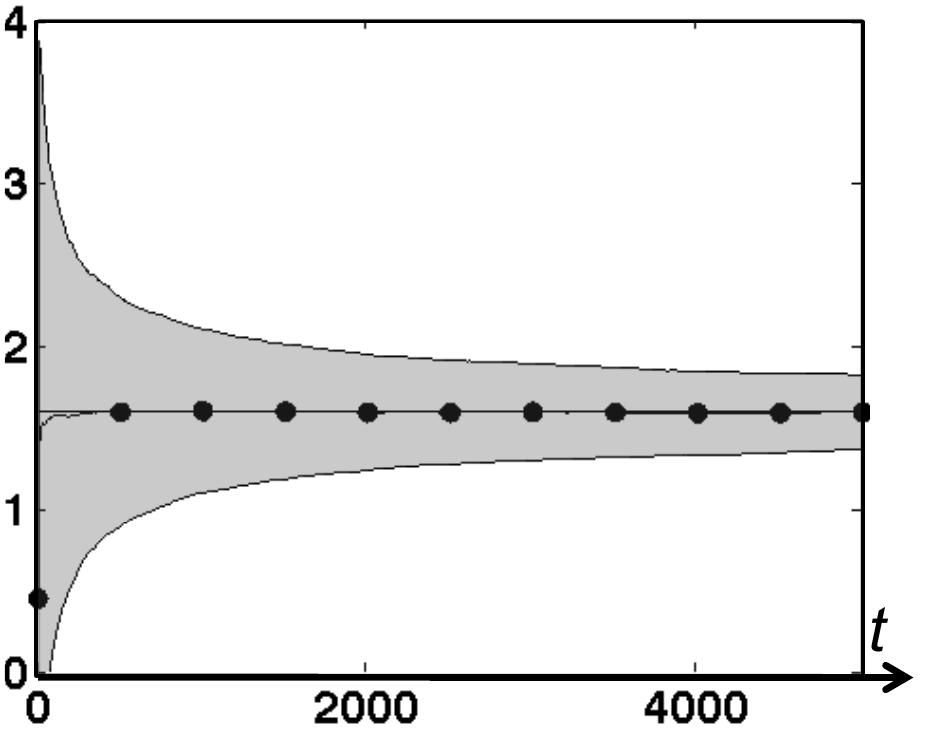}}
 \subfigure[]{ \includegraphics[width=0.3\textwidth]{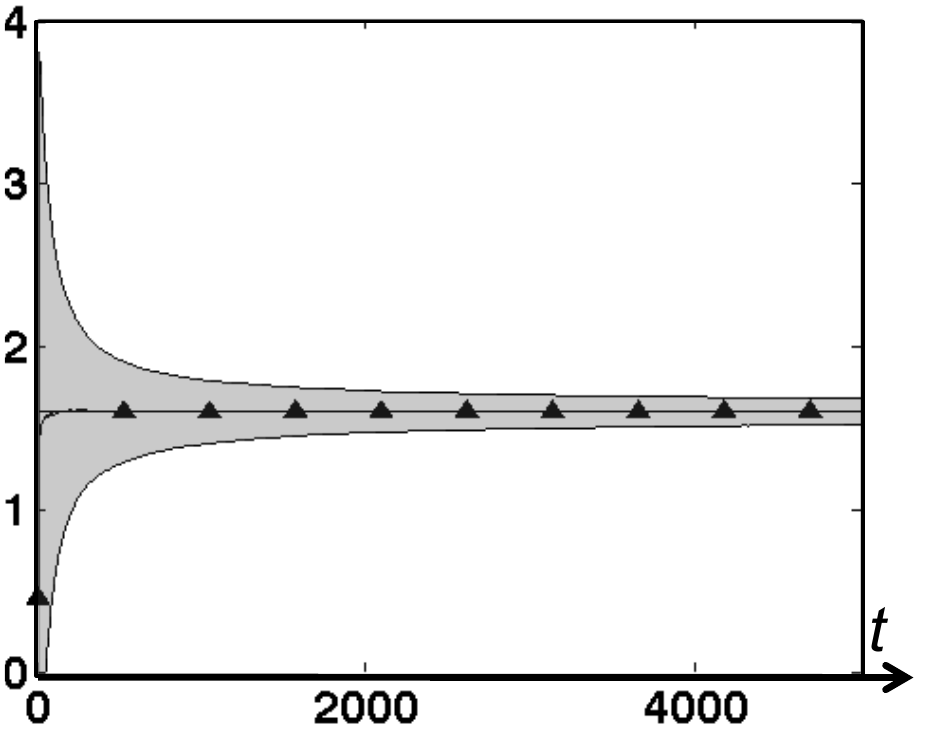}}
 \subfigure[]{\includegraphics[width=0.3\textwidth]{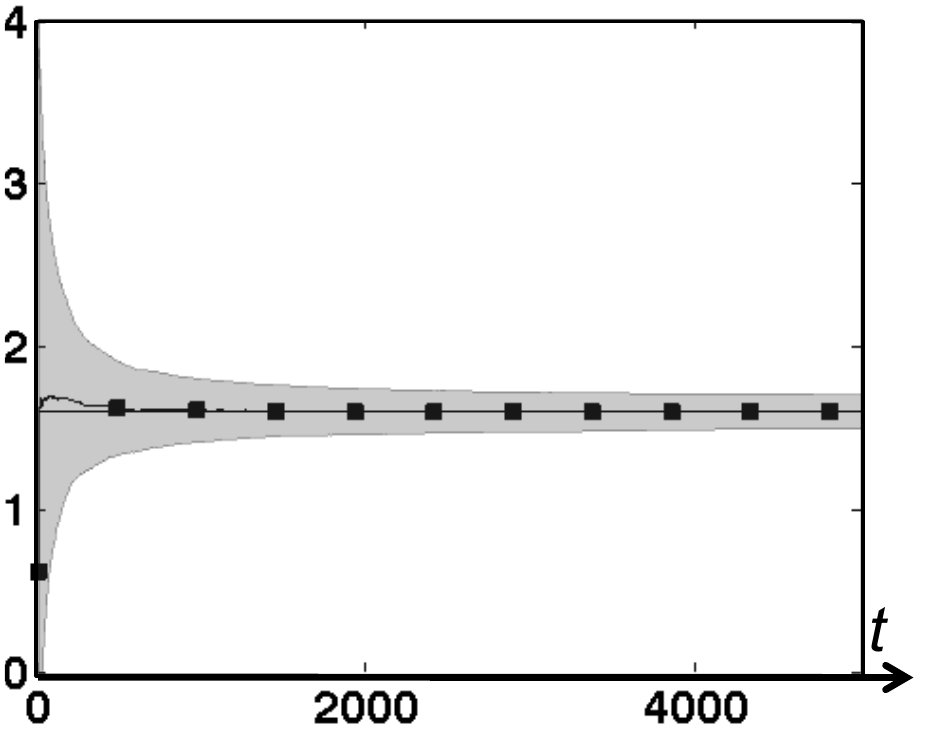}}
}
\centerline{
\subfigure[]{\includegraphics[width=0.3\textwidth]{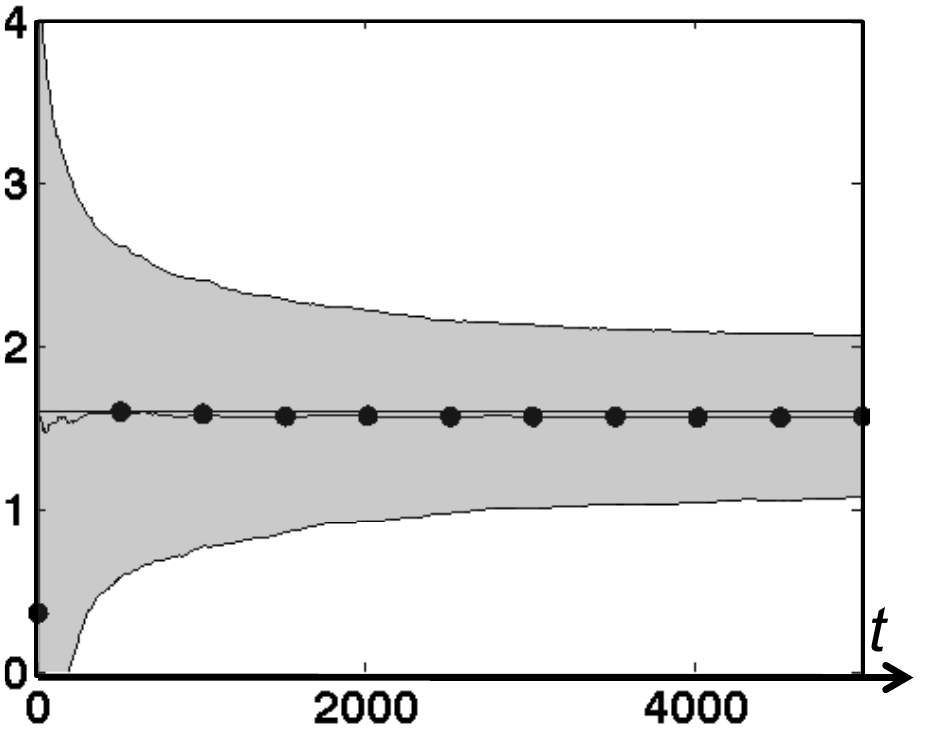}}
 \subfigure[]{ \includegraphics[width=0.3\textwidth]{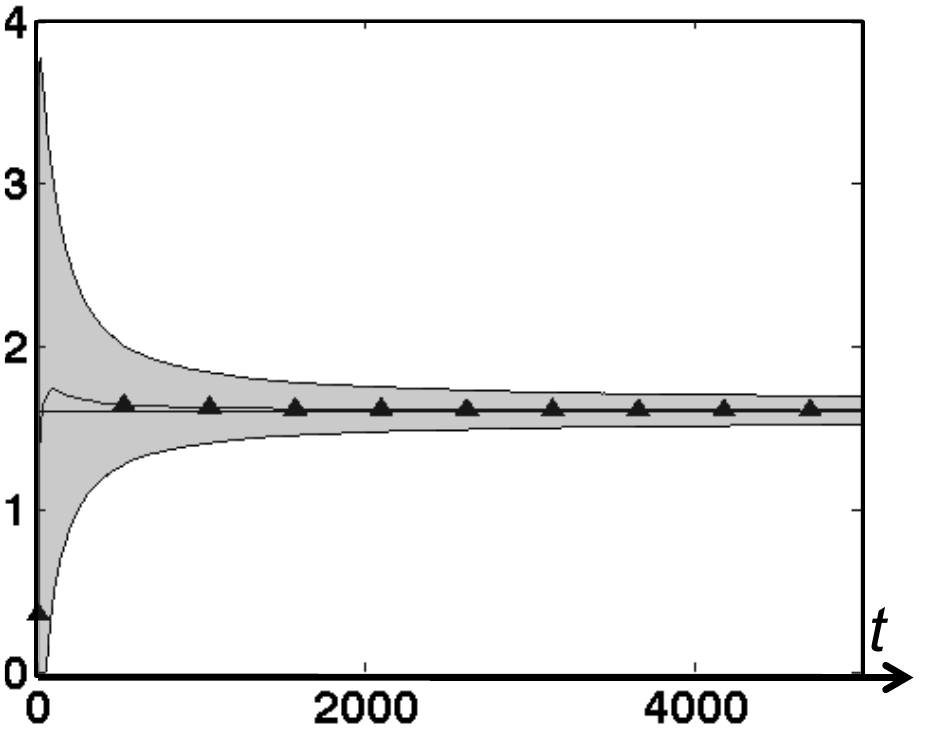}}
 \subfigure[]{\includegraphics[width=0.3\textwidth]{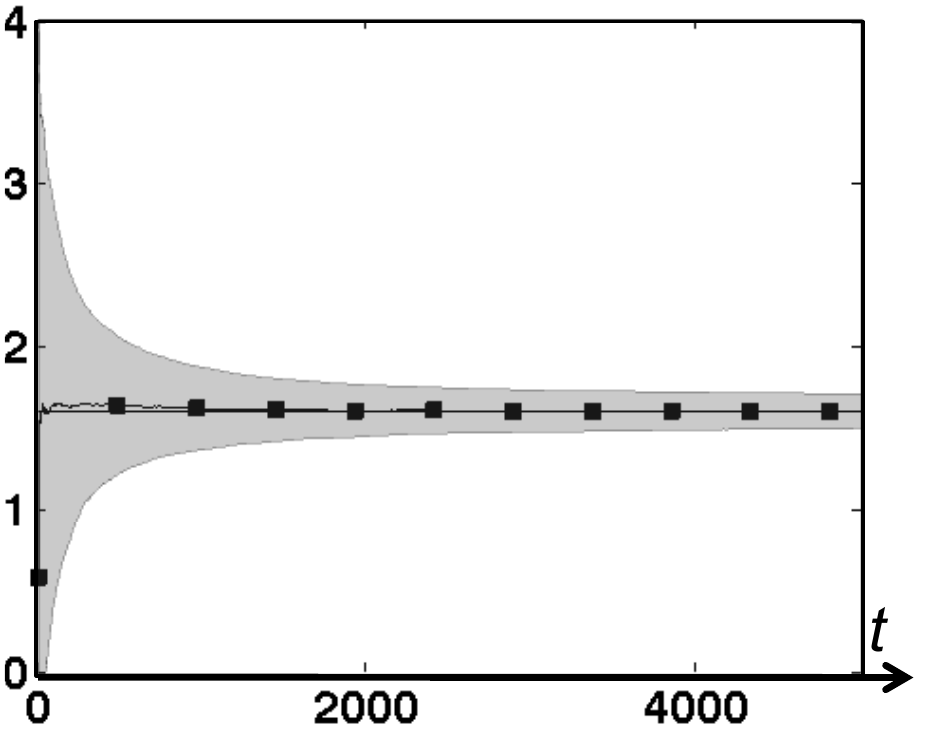}}
}
  \caption{Estimated mean as function of the iteration index $t$, averaged over $2000$ runs, for the ARMS (Figures  {\bf (a)}, {\bf (d)}, {\bf (g)} and {\bf (j)}), A$^2$RMS (Figures {\bf (b)}, {\bf (e)}, {\bf (h)} and {\bf (k)}) and IA$^2$RMS (Figures {\bf (c)}, {\bf (f)}, {\bf (i)} and {\bf (l)}). Figures {\bf (a)}, {\bf (b)} and {\bf (c)} correspond to $\pi_t(x)$ obtained using Eqs. (\ref{EqWtARMS})-(\ref{EqWtARMS2}). Figures {\bf (d)}, {\bf (e)} and {\bf (f)} to $\pi_t(x)$ given by (\ref{EqSimpler1}). Figures {\bf (g)}, {\bf (h)} and {\bf (i)} to $\pi_t(x)$ given by (\ref{EqSimpler2}). Figures {\bf (j)}, {\bf (k)} and {\bf (l)} are obtained using the procedure described in Eq. (\ref{EqSimpler3}).
}
\label{figResultsA}
\end{figure}

First of all, from these three tables we note that the two proposed algorithms provide better results than the standard ARMS in all cases, regardless of the scheme used to build the proposal. This can be seen by the slight improvement in the estimated mean averaged over the $2,000$ runs (we recall that the true mean is $1.6$) and, most notably, by the large decrease in the standard deviation of the estimation. This can also be clearly appreciated in Figure \ref{figResultsA}, where the average value of the estimated mean, plus and minus the average standard deviation, is plotted for $t=0,\ 1,\ \ldots,\ 5000$. Although all the curves converge to the true mean, the ones obtained using the two variants of ARMS proposed (A$^2$RMS and IA$^2$RMS) show clearly a reduced variance in the estimation, especially after the $5000$ iterations have been performed.

 
Regarding the comparison of ARMS with the two proposed algorithms performed in Tables {\ref{resultsARMS}, \ref{resultsA2RMS} and \ref{resultsIA2RMS}}, we also notice that the correlation after $N=5000$ iterations is much lower for the two variants of ARMS proposed than for the standard ARMS. Furthermore, the convergence of $\pi_t(x)$ to $p(x)$ also improves greatly, as evidenced by the value of $D_{\pi|p}(t)$, which is more than one order of magnitude lower than for the standard ARMS after $N=5000$ iterations, both for A$^2$RMS and IA$^2$RMS.

We remark that all these improvements come at the expense of a larger number of support points (whose amount remains moderate anyway, not increasing without bound as $t \to \infty$), and only at a slight increase in the computational cost, since both variants are very similar to ARMS, inserting just an additional step to incorporate support points inside regions where $\pi_t(x) < p(x)$.

We also notice that the third procedure for building the proposal pdf, described by Eq. \eqref{EqSimpler2}, always provides the best results (regarding the average mean and standard deviation), even better than the first procedure with the standard ARMS, which is the strategy proposed in the original paper. 
Indeed, the third procedure provides a good approximation of all the modes of the target pdf {\it more quickly} than the other procedures, obtained at the cost of using a larger number of support points. 
  This is owing to the procedure 3 having a greater ``disposition'' to propose and then incorporate support points in different regions of the domain, since it is formed by pieces of uniform pdfs.
However, stepped lines are not the best approximation of a curve (in this case, the shape of $p(x)$), hence the  distance $D_{\pi|p}(x)$ with the procedure 3 decreases more slowly than the rest. 

The second procedure attains a good approximation of the target $p(x)$ ($D_{\pi|p}(x)$ close to zero) always using the smallest number of support points. However, it provides the worst results in terms of estimation. 
Finally, the fourth procedure arguably provides the best trade-off among computational cost, approximation of the target density and accuracy of the estimation. 
\section{Discussion}
\label{conclusion_Sect}

The efficiency of MCMC methods depends on the similarity of the proposal and the target densities.
In fact, the relationship between the speed of convergence of MCMC approaches and the discrepancy between the proposal and the target pdf has been remarked in different works (see e.g. \citep{Holden98}).
Hence, in order to minimize this discrepancy and speed up the convergence, many adaptive strategies have been proposed in the literature \citep{Andrieu06,Andrieu08,Cai08,Gasemyr03,Griffin11,Haario01,Holden09}.
Indeed, the {\it automatic} construction of a sequence of proposal pdfs that are both easy to draw from and good approximations to the target density, is a very important issue (and a difficult task) to speed up MCMC algorithms. 

In order to draw from complicated univariate target densities, one popular approach for building the proposal adaptively is adaptive rejection Metropolis sampling (ARMS), which is an important generalization of standard adaptive rejection sampling (ARS).
ARMS was introduced to allow drawing samples from univariate non-log-concave distributions through the addition of a Metropolis-Hastings step, and reduces to ARS for log-concave target pdfs.
Furthermore, it is important to remark that ARMS is often used to handle {\it multivariate} distributions within MCMC methods that require sampling from univariate full conditional distributions, such as Gibbs sampling \citep{Robert04}, the hit-and-run algorithm \citep{Belisle93} and adaptive direction sampling \citep{Gilks94}.

In this work, we have first highlighted that the approach followed by ARMS to build the proposal can be decoupled from the procedure used to incorporate points to the support set. Moreover, we have pointed out a structural limitation of ARMS (support points can never be added inside regions where $\pi_t(x) < p(x)$) that may prevent the target from converging to the proposal.

In order to overcome this drawback, we have proposed two alternative ARMS-type adaptive schemes that guarantee both the convergence of the proposal to the target density and the convergence of the Markov chain to the invariant distribution, including the statistical information provided by every sample $ x'\in \mathcal{D}$ such that $\pi_t(x') < p(x')$.
The first one (A$^2$RMS) is a direct modification of the ARMS procedure that incorporates support points inside regions where the proposal is below the target with a decreasing probability, thus satisfying the diminishing adaptation property, one of the required conditions to assure the convergence of the Markov chain.
The second one (IA$^2$RMS) is an adaptive independent MH algorithm with the ability to learn from all previous samples except for the current state of the chain, thus also ensuring the convergence of the chain to the invariant density \citep{Gasemyr03,Holden09}, \cite[Chapter 8]{Liang10}.
The underlying idea behind both of the proposed schemes is measuring the discrepancy between the proposal and target densities through the computation of ratios of the two pdfs.  


We have shown through a numerical example that the two novel approaches proposed provide an improvement in performance w.r.t. standard ARMS: more accurate estimated mean, less variance in the estimation, less correlation between consecutive samples and better convergence to the target pdf.
Regarding this last issue, it is important to remark that the two ARMS-type procedures introduced guarantee that the discrepancy between the proposal and the target pdf decreases quickly, implying that the proposal becomes closer and closer to the target density.
Note that this is not ensured in other adaptive MH algorithms \citep{Haario01,Liang10}, which simply optimize certain parameters of a proposal pdf with a fixed analytical form, whereas here we improve the entire shape of the proposal density, ensuring the convergence to the target pdf.

The price paid w.r.t. standard ARMS is a slight increase in computational cost, due to the single additional step to allow for the incorporation of support points inside regions where $\pi_t(x) < p(x)$, and a moderate increase in the number of support points, which remains bounded anyway. Moreover, the effort to code and implement the novel algorithms remains virtually unchanged, since the new control step in A$^2$RMS and IA$^2$RMS does not need additional evaluations of the proposal and the target pdfs.

Finally, we have also noticed that these novel techniques allow us to reduce the complexity in the construction of the sequence of proposals w.r.t. standard ARMS, thus compensating the slight increase in computational cost.
Consequently, we have also proposed different simpler procedures to build the sequence of proposal densities and tested them with the standard ARMS, A$^2$RMS and IA$^2$RMS. This is an important issue as evidenced, for instance, by the fact that the work of \citep{Meyer08} is dedicated exclusively to an alternative construction of the proposal pdf within the ARMS technique.

One of introduced procedures, constructed using pieces of uniform pdfs plus two exponential-type pdfs for the tails, is particularly interesting due to its simplicity and good performance even with standard ARMS. Indeed, it could potentially be used to design efficient ARMS algorithms to draw {\it directly} from multidimensional target distributions. 
Moreover, the procedure based on the work of \citep{Cai08} that uses a piecewise linear approximation of the target pdf and two exponential tails, provides the best trade-off among computational cost, approximation of the target density and accuracy of the estimation.






\section{Acknowledgment}
This work has been partly financed by the Spanish government, through the DEIPRO project (TEC2009-14504-C02-01) and the CONSOLIDER-INGENIO 2010 Program (ProjectCSD2008-00010).

The authors would also like to thank Roberto Casarin (Universit\'a Ca' Foscari di Venezia) and Fabrizio Leisen (Universidad Carlos III de Madrid) for many useful comments and discussions about the ARMS technique.


\bibliography{bibliografia} 
   
\end{document}